\documentclass[prb, twocolumn,superscriptaddress, preprintnumbers, letterpaper]{revtex4-1}

\usepackage{times}
\usepackage{amsmath}
\usepackage{amsthm}
\usepackage{amssymb}
\usepackage{amsbsy}
\usepackage{amsfonts}
\usepackage{graphicx}
\usepackage{bibunits}
\usepackage{resizegather}
\usepackage{xr}
\usepackage{tocvsec2}

\makeatletter
\let\O\@undefined
\def\O{\mathrm{O}}

\newcommand{\nc}{\newcommand}
\nc{\CA}{{\cal A}} \nc{\CB}{{\cal B}} \nc{\CC}{{\cal C}}
\nc{\CD}{{\cal D}} \nc{\CE}{{e}} \nc{\CF}{{\cal F}}
\nc{\CG}{{\cal G}} \nc{\CH}{{\cal H}} \nc{\CI}{{\cal I}}
\nc{\CJ}{{\cal J}} \nc{\CK}{{\cal K}} \nc{\CL}{{\cal L}}
\nc{\CM}{{\cal M}} \nc{\CN}{{\cal N}} \nc{\CO}{{\cal O}}
\nc{\CP}{{\cal P}} \nc{\CQ}{{\cal Q}} \nc{\CR}{{\cal R}}
\nc{\CS}{{\cal S}} \nc{\CT}{{T}} \nc{\CU}{{\cal U}}
\nc{\CV}{{\cal V}} \nc{\CW}{{\cal W}} \nc{\CX}{{\cal X}}
\nc{\CY}{{\cal Y}} \nc{\CZ}{{\cal Z}}

\nc{\bA}{\mathbb{A}} \nc{\bB}{\mathbb{B}} \nc{\bC}{\mathbb{C}}
\nc{\bD}{\mathbb{D}} \nc{\bE}{\mathbb{E}} \nc{\bF}{\mathbb{F}}
\nc{\bG}{\mathbb{G}} \nc{\bH}{\mathbb{H}} \nc{\bI}{\mathbb{I}}
\nc{\bJ}{\mathbb{J}} \nc{\bK}{\mathbb{K}} \nc{\bL}{\mathbb{L}}
\nc{\bM}{\mathbb{M}} \nc{\bN}{\mathbb{N}} \nc{\bO}{\mathbb{O}}
\nc{\bP}{\mathbb{P}} \nc{\bQ}{\mathbb{Q}} \nc{\bR}{\mathbb{R}}
\nc{\bS}{\mathbb{S}} \nc{\bT}{\mathbb{T}} \nc{\bU}{\mathbb{U}}
\nc{\bV}{\mathbb{V}} \nc{\bW}{\mathbb{W}} \nc{\bX}{\mathbb{X}}
\nc{\bZ}{\mathbb{Z}}

\nc{\sgn}{\, {\rm sign}}

\begin{document}


\title{Transport in a One-Dimensional Hyperconductor}

\author{Eugeniu Plamadeala}
\affiliation{Department of Physics, University of California, Santa Barbara,
California 93106, USA}
\author{Michael Mulligan}
\affiliation{Stanford Institute for Theoretical Physics, Stanford 
University, Stanford, CA 94305, USA}
\author{Chetan Nayak}
\affiliation{Department of Physics, University of California, Santa Barbara, California 93106, USA}
\affiliation{Microsoft Research, Station Q, Elings Hall,
University of California, Santa Barbara, California 93106-6105, USA}

\begin{abstract}
We define a `hyperconductor' to be a material whose
electrical and thermal DC conductivities are infinite at zero temperature and finite at any non-zero temperature.
The low-temperature behavior of a hyperconductor is controlled by a quantum critical 
phase of interacting electrons that is stable to all potentially-gap-generating interactions and potentially-localizing disorder. 
In this paper, we compute the low-temperature DC and AC electrical and thermal conductivities in
a one-dimensional hyperconductor, studied previously by the present authors,
in the presence of both disorder and umklapp scattering. 
We identify the conditions under which the transport coefficients are finite, which allows us to exhibit examples of violations of the Wiedemann-Franz law. 
The temperature dependence of the electrical conductivity, which is characterized by the parameter $\Delta_X$, is a power law, $\sigma \propto 1/T^{1 - 2(2-\Delta_X)}$ when $\Delta_X \geq 2$, down to zero temperature when the Fermi surface is commensurate with the lattice. There is a surface
in parameter space along which $\Delta_X = 2$ and $\Delta_X \approx 2$ for small deviations from this surface.
In the generic (incommensurate) case with weak disorder, such scaling is seen at high-temperatures,
followed by an exponential increase of the conductivity $\ln \sigma \sim 1/T$
at intermediate temperatures and, finally, $\sigma \propto 1/T^{2-2(2-{\Delta_X})}$ at the lowest temperatures.
In both cases, the thermal conductivity diverges at low temperatures.
\end{abstract}

\maketitle

\tableofcontents

\section{Introduction}

\subsection{Goal of this paper}

In this paper, we study transport in the one-dimensional non-Fermi liquid introduced in Ref. [\onlinecite{Plamadeala2014}]. This metallic phase is very different from a Fermi liquid:
in addition to anomalous single-electron properties,
it is a perfect metal at zero-temperature, with infinite DC conductivity even in the presence of impurities,
unlike a Fermi liquid. We call such a material a ``hyperconductor,'' to distinguish it from a superconductor,
since a hyperconductor does not have a Meissner effect at zero temperature; its electrical conductivity is finite
at any non-zero temperature; and its thermal conductivity diverges as the temperature approaches zero.
The goal of this paper is to compute the temperature and frequency dependence
of the electrical and thermal conductivity of a hyperconductor at low temperature.
The temperature dependence of the conductivities is characterized by the parameter $\Delta_X$ and depends on whether the Fermi
surface is commensurate with the lattice.
In the commensurate case, {\it both} the electrical $\sigma$ and thermal $\kappa$ conductivities behave as a power law: $\sigma, \kappa \propto 1/T^{1 -  2 (2-\Delta_X)}$ with the special case $\Delta_X=2$
occurring along a surface in parameter space.
This constitutes a violation of the Wiedemann-Franz ``law," which states that the ratio $\kappa/\sigma T$ is constant, and is due to differing relaxation mechanisms of the electrical and thermal currents.
In the incommensurate case, there is a range of temperatures over which both $\sigma$ and $\kappa$
diverge exponentially, although with differing algebraic prefactors, as $T \rightarrow 0$; at the lowest temperatures, $\sigma \propto \kappa/T \propto 1/T^{2 - 2 (2-\Delta_X)}$.
The above temperature dependences reflect the non-Fermi liquid physics of this hyperconductor.
As a concrete and well-controlled example of transport in a non-Fermi liquid, these results
may shine light on general principles regarding non-Fermi liquids and transport in strongly-correlated
electron systems.

\subsection{General remarks about metallic transport}
\label{sec:remarks}

Transport provides one of the most important characterizations of a physical system.
It is often said that the DC electrical conductivity is the first property to be measured when a new material is investigated.
However, this is usually followed by noting that it is often the last property to be understood, highlighting the subtle nature of
transport properties, when compared with thermodynamic ones.\cite{Kivelsonlecture2011} 
This is one of the difficulties involved in understanding metallic
states whose low-temperature behavior is not controlled by the Fermi liquid fixed point
but by some other fixed point -- generally called a `non-Fermi liquid'. 
Experimental systems that are candidate non-Fermi liquid metals have primarily been identified by the occurrence
of DC conductivity exhibiting unusual temperature dependence. Perhaps the most famous example is the normal state of the cuprate high-temperature superconductors\cite{Carlson2002, Lee2006} around optimal doping, where the DC electrical conductivity $\sigma \sim 1/T$ over a large range of temperatures $T$.
It is difficult to construct models that show such behavior; non-Fermi liquids\cite{Varmaphenomenlogy1989, AndersonLuttinger1990, halperinleeread1993, AltshulerIoffeMillis1994, Polchinski1994, NayakWilczek1994short, NayakWilczeklong1994, ChakravartyNortonSyljuasen1995, Dobrosavliev97, Chakravarty98, OganesyanKivelsonFradkin2001, MetznerRoheAndergassen2003, AbanovChubukov2004, Lawleretal2006, SSLee2009OrderofLimits, Mross2010, MetlitskiSachdev2010Part1, MetlitskiSachdev2010Part2, HartnollHofmanMetlitskiSachdev2011, FaulknerIqbalMcGreevyVegh2011, StanfordGroup2013, StanfordGroup2014} (e.g., fermion-gauge field systems) often have more pronounced anomalies in single-particle properties, but more conventional behavior in transport.\cite{KimFurusakiWenLee1994}
See Refs. [\onlinecite{FaulknerIqbalLiuMcGreevyVegh2010}] and [\onlinecite{HartnollMahajanPunkSachdev2014}]
for two counterexamples.

The rate at which the conductivity of a metal approaches its zero-temperature value is determined by the available relaxation mechanisms,
which are, in turn, reflective of the nature of the zero-temperature metallic state.
In a clean Fermi liquid, umklapp scattering provides the leading low-temperature momentum-relaxation mechanism and results in the familiar contribution, $\delta \rho_{xx}(T) \propto T^2$, in spatial dimensions $D>1$, \cite{zimanbook, AshcroftMermin} to the DC electrical resistivity.
\footnote{In $D=1$, the umklapp scattering may result in a linear dependence upon temperature [\onlinecite{GiamarchiMillis92}].}
In 3D, an electron-phonon interaction contributes $\delta \rho_{xx} \propto T^5$ below the Debye temperature,
while $\rho_{xx}(T) \propto T$ is found above the Debye temperature.\cite{AshcroftMermin} 
Similar behavior is found for the
scattering of electrons by other collective bosonic modes. 
However, at the lowest temperatures, which is inevitably below the Debye temperature or its analogues
for other collective bosonic modes, the resistivity vanishes faster than linearly in almost all theoretical models. 

One way to understand this is as follows.
In a metal, the resistivity generally vanishes at low temperatures as $\rho \sim 1/\tau_{\rm tr}$, where $\tau_{\rm tr}$ is the decay rate for the current, usually called
the transport lifetime. 
On dimensional grounds, $1/\tau_{\rm tr} \propto {(g T^{-\Delta_g})^2} \cdot T$ where $g$ is the coupling constant that dominates the relaxation of the current and $\Delta_g$ is its scaling dimension. 
(For umklapp-dominated relaxation,
$g$ is the strength of umklapp scattering process and $\Delta_g$ is its scaling dimension, with ${\Delta_g}=2-\Delta_{X}$
if $X$ is the umklapp scattering operator specified in Eq. \ref{umklappscattering}. For disorder-dominated relaxation,
$g^2$ is the variance of the disorder and $2\Delta_g$ is its scaling dimension,
with $2\Delta_g = 3-2\Delta_X$ if $X$ is the operator that is coupled to disorder
in Eq. \ref{disorderscattering}.) 
If the coupling $g$ is an irrelevant perturbation, $\Delta_g < 0$, (including the
case of a marginally irrelevant perturbation) at the
zero-temperature metallic fixed point, then the resistivity vanishes faster than linearly with $T$, which is the usual case. 
If, on the other hand, $g$ is
a relevant or marginally relevant perturbation, $\Delta_g > 0$, then the fixed point is not stable,
and the ultimate low-temperature behavior is determined by some other fixed point. 
Hence,
$\rho \propto T$ can only occur in a model that contains a strictly marginal operator, $\Delta_g = 0$, that relaxes the current.
This, in turn, implies that an observed $\rho \propto T$ is either an intermediate temperature behavior that does
not survive to the lowest of temperatures, as in the case of
electron-phonon scattering above the Debye temperature, or it is a consequence of physical processes encapsulated by a strictly marginal operator.
See Refs. [\onlinecite{PhillipsChamon2005, Hartnoll2010, Hartnoll2015}] for related scaling arguments.

The $23$-channel Luttinger liquid parameter regime that was called the
`asymmetric shorter Leech liquid' in Ref. [\onlinecite{Plamadeala2014}]
has many such marginal operators. This model is a 1D hyperconductor, in the sense defined above: its
electrical and thermal conductivities diverge at zero temperature in the presence of arbitrary (perturbative) electron-electron and disorder-mediated interactions. However, the temperature and frequency
dependence of these transport coefficients is interesting because of the presence of these marginal operators.
The purpose of this paper is to explore this dependence.

In the presence of conservation laws, there is an important caveat to the scaling considerations given above.\cite{Giamarchi1991,GiamarchiMillis92,  ZotosNaefPrelovsek1997, RoschAndrei2000, Shimshoni2003, SirkerPereiraAffleck2011, KarraschBardarsonMoore2012}
Some theoretical models may have conservation laws that prevent the electrical and/or thermal
currents from fully relaxing, thereby leading to infinite conductivities. Some care is required in these
cases, since approximate calculations of transport relaxation times $\tau_{\rm tr}$ may give
finite answers due to the failure of these approximations to properly account for these conservation laws.
An additional complication is that the Fermi momentum $k_F$ and the reciprocal lattice vectors ${\bf G}$ enter into (pseudo)-momentum conservation for low-energy excitations.
As a result, these momentum scales, which are nominally
short-distance or ultraviolet scales, may enter into the low-temperature, low-frequency response.\cite{NayakWilczekPhysProp1996}
Conservation laws, together with these momentum scales, may conspire to modify
the simple scaling form $1/\tau_{\rm tr} \propto {(gT^{- \Delta_g})^2} \cdot T$ to
$1/\tau_{\rm tr} \propto {(g T^{- \Delta_g})^2} \cdot T \cdot f(p/T)$, where $f(x)$ is a scaling function that could have, for instance, the asymptotic form $f(x)\sim e^{-x}$ for large $x$ and $p$ is some characteristic momentum
(e.g. a combination of the Fermi momentum and reciprocal lattice vectors) that is relevant to the relaxation of the current.
One possible consequence is that the Wiedemann-Franz law may be implied by scaling,
but need not be realized because of symmetry considerations.

\subsection{Organization of this paper}

The remainder of this paper is organized as follows.
In Sec. \ref{PMreview}, we review the construction of the hyperconductor of Ref. [\onlinecite{Plamadeala2014}].
In Sec. \ref{symmetryandtransport}, we discuss the relation between conservation laws and dissipative transport with an eye towards the application to the hyperconductor phases.
In Sec. \ref{PMtransport}, we calculate the electrical and thermal conductivities of the hyperconductor at both commensurate and incommensurate filling for a pure system with umklapp scattering and a weakly disordered system.
The memory matrix formalism provides the calculational tool of this section.
We conclude and outline future plans in Sec. \ref{Conclusion}.
We include three appendices that provide details for the calculations underlying the results presented in Sec. \ref{PMtransport}.

\section{Review of the 1D Hyperconductor}
\label{PMreview}

In this section, we give a highly condensed review of the derivation of the hyperconductor
of Ref. [\onlinecite{Plamadeala2014}] in order to establish notation that is used in the remainder of this paper.
For the most part in this paper, when we use the term, hyperconductor, we specifically have in mind the example previously called the 1D `asymmetric shorter Leech liquid,' however, we emphasize that the notion is more general and we are merely studying one particular realization.
The reader interested in the details of this construction is directed to Ref. [\onlinecite{Plamadeala2014}].

The 1D hyperconductor that is the subject of this paper obtains from the low-energy effective theory of a particular interacting model of electrons in a 1D quantum wire. We can regard the bands with different values of the transverse momentum, as well as the two spin states of the electron, as separate channels. The simplest example then, and the one we will study in this paper has $N=23$ channels of spinless fermions $\Psi_I$. 

At low energies, the non-relativistic fermions can be linearized into a theory of $N=23$ channels of chiral linearly-dispersing spinless (Dirac) fermions, with a left and a right mover in each channel. Their complete action is given by:
\begin{equation}
\label{relativistic}
S_{\rm lin} =  S_0 + S_\text{int}
\end{equation}
\begin{multline}
S_0 =  \int_{t,x}  \Big[\psi^\dagger_{R,I} i(\partial_t + {v_I}\partial_x) \psi_{R,I} \\ +
\psi^\dagger_{L,I} i (\partial_t - {v_I}\partial_x)\psi_{L,I}\Big]
\end{multline}
\begin{multline}
S_\text{int} =  \int_{t,x}  \Big[ U_{I, J} \psi^\dagger_{R, I} \psi_{R, I} \psi^\dagger_{R, J} \psi_{R, J} \\
 + U_{I+N, J+N} \psi^\dagger_{L, I} \psi_{L, I} \psi^\dagger_{L, J} \psi_{L, J}\\ 
 + 2U_{I, J+N} \psi^\dagger_{R, I} \psi_{R, I} \psi^\dagger_{L, J} \psi_{L, J} \Big]
\end{multline}
where the operator $\psi^\dagger_{R,I}$ ($\psi^\dagger_{L,I}$) creates a right-moving (left-moving) fermion excitation about the Fermi point $k_{F,I}$ ($-k_{F,I}$) in channel $I=1,\ldots,N$ and we have used the short-hand $\int_{t,x} \equiv \int dt dx$.
The velocity of the $I^\text{th}$ channel of fermions is $v_I$.
It is important to keep in mind that the linear regime only includes momenta smaller than some cutoff $\Lambda$, where $\Lambda \ll k_F$

As the real symmetric matrix $U_{I,J}$ for $I, J = 1, \ldots, 2N$ specifying the density-density interaction is varied, the system explores the parameter space of a $23$-channel Luttinger liquid. 
As discussed in Ref. [\onlinecite{Plamadeala2014}], there is an open set of $U_{I,J}$
for which all potentially-gap-opening or potentially-localizing perturbations to Eq. \ref{relativistic} are irrelevant;
this entire parameter regime is the hyperconductor phase.
The calculations of Ref. [\onlinecite{Plamadeala2014}] that establish the existence of
this phase as well as the following transport calculations rely on the
bosonic representation of Eq. (\ref{relativistic}):
\begin{equation}
\label{eqn:LL-action}
    S_{\rm b} = \frac{1}{4\pi} \int_{t,x} \Big[ K_{IJ} \partial_t \phi_I \partial_x \phi_J - V_{IJ} \partial_x \phi_I \partial_x \phi_J \Big].
\end{equation}
with $K=K_{\rm{ferm}}= - \mathbb{I}_N\oplus \mathbb{I}_N$, $V_{IJ}={v_I}\delta_{IJ} + U_{IJ}$, $\mathbb{I}_N$ the $N\times N$ identity matrix, and $I, J = 1, \ldots, 2N$ in Eq. (\ref{eqn:LL-action}).
The operators $\psi^\dagger_{I,R}=\frac{1}{\sqrt{2\pi a}}e^{i\phi_{I}}\gamma_{I}$ and $\psi^\dagger_{I,L}=\frac{1}{\sqrt{2\pi{a}}}e^{-i\phi_{I+N}}\gamma_{I+N}$ create, respectively, right- and left-moving fermions in the $I^{\rm{th}}$ channel; $a$ is a short-distance cutoff, and the Klein factors $\gamma_I$ satisfy $\gamma_J\gamma_K=-\gamma_K\gamma_J$ for $J\neq K$. The bosonic fields satisfy the equal-time commutation relations $ \left[ \phi_I(x), \Pi_J(y) \right] = i \delta_{I,J} \delta(x-y)$, where the canonical momenta $\Pi_I = \frac{1}{2\pi} K_{IJ} \partial_x \phi_J$.
(The index on the fields $\Psi_{I, R/L}$ runs from $1, \ldots, N$, while the index on the bosonic fields $\phi_I$ runs from $1, \ldots 2N$.)

The hyperconductor construction is based on the observation that under an $SL(2N,\mathbb{Z})$ basis change, 
$\phi_I \equiv W_{IJ}\tilde{\phi}_J$, it is possible to transform $K$ to the Gram matrix $\tilde{K} = W^T K W = - \tilde{K}_R \oplus \tilde{K}_L$ of a signature $(N,N)$ lattice of the form $- \tilde{\Lambda}_R \oplus \tilde{\Lambda}_L$ where $\tilde{\Lambda}_R$, $\tilde{\Lambda}_L$ are positive-definite
unimodular\footnote{The unimodularity follows from the unimodularity of $K_{\rm{ferm}}= - \mathbb{I}_N\oplus \mathbb{I}_N$ and the determinant-preserving property of $W$.} $N$-dimensional lattices. 
For $N\geq 23$, there exist {\it non-root} positive-definite unimodular lattices -- i.e., lattices
such that all vectors ${\bf v}$ in the lattice satisfy $|{\bf v}|^2 > 2$ -- and there
exist matrices $W$ that transform $K_{\rm{ferm}}$ to the corresponding Gram matrices.
If, in this basis, $\tilde{V} = W^T V W$ is block diagonal (i.e., does not mix right-movers and left-movers), then all potentially gap-opening or localizing operators $\cos(\tilde{m}_I \tilde{\phi}_I)$
are irrelevant when $\tilde{\Lambda}_R$ {\it or} $\tilde{\Lambda}_L$ is non-root, where $\tilde{m}_J = m_I W_{IJ}$.
Stability persists for a small but finite range of values of any parameters in the model
(i.e., away from block diagonal $\tilde{V}$), including the chemical potentials in each channel,
the velocities, and all the inter-channel and inter-spin interactions. In the hyperconductor phase considered
in this paper, $\tilde{\Lambda}_R$ is the so-called shorter Leech lattice, the unique non-root
unimodular integral lattice in $23$ dimensions, while $\tilde{\Lambda}_L$ is $\mathbb{Z}^{23}$, the ordinary hypercubic lattice, which is {\it not} a non-root lattice. 
This phase was called the {\it asymmetric shorter Leech liquid}.
(See Refs. \onlinecite{PlamadealaE8, generalstableequivalence} for a fuller discussion of the mathematical technology underlying the hyperconductor construction.)

For simplicity, we perform the calculations in this paper using an
interaction matrix $\tilde{V}_{IJ}$ in the transformed basis that
is simply proportional to the positive-defined matrix $\tilde{K}_R \oplus \tilde{K}_L$, so that all of the eigenmodes have equal velocities $v$.
We similarly assume, for simplicity, that $k_{F,I}=k_F$ for all $I$.

The salient feature of the asymmetric shorter Leech hyperconductor that is relevant to this paper is the existence of a large number of marginal backscattering operators of the form $\cos\left(\tilde{m}_I \tilde{\phi}_I\right)$
when $\tilde{V} = W^T V W$ is block diagonal and $\tilde{\Lambda}_R$
and $\tilde{\Lambda}_L$ are, respectively, the shorter Leech lattice and $\mathbb{Z}^{23}$.
In conformal field theory\cite{ginspargappliedcft1991} (CFT) terminology, these operators have different right and left scaling dimensions $\left (\Delta_R, \Delta_L \right ) = \left (\frac{3}{2}, \frac{1}{2} \right )$.
If $\tilde{V}$ is moved slightly away from block diagonal, then the scaling dimensions of
any such operator will be shifted to $\left (\Delta_R, \Delta_L \right ) = \left (\frac{3}{2}+y, \frac{1}{2} + y\right )$, where $y$ will
depend on the particular operator in question.
For block diagonal $\tilde{V}$, these scaling dimensions are protected by their
chirality: their RG equations do not contain higher-order terms.\cite{Cardypotts1993}
(See Appendix \ref{marginalityappendix} for a review of this argument.)
As a result, transport coefficients exhibit anomalous power-law dependence all the way
to zero temperature. For block diagonal $\tilde{V}$, this is manifested as DC electrical resistivity
$\rho_{\rm DC} \propto T$ all the way to zero temperature.

\section{Symmetry and Transport}
\label{symmetryandtransport}

In this section, we describe some of the complications associated with computing the transport
properties of a 23-channel Luttinger liquid. Most of the material in this section has been
described elsewhere (see below for references) but, for the sake of completeness,
we give a review of transport that is tailored to the application of the formalism described in the next section.
The reader that is interested primarily in our results may wish to skip this rather technical section
on a first reading of this paper.

\subsection{Conservation Laws}

The conservation of total electrical charge and total energy,
\begin{equation}
\frac{d}{dt} Q = \frac{d}{dt} H = 0,
\end{equation}
(where $Q$ and $H$ are the total electrical charge and energy operators) make it possible for those quantities to diffuse, thereby leading to finite electrical
and thermal conductivities. If, however, the charge or energy {\it currents}, respectively
${\bf J}^e$ or ${\bf J}^T$, were conserved,
\begin{equation}
\frac{d}{dt} {\bf J}^e = 0 \hskip 0.5 cm \mbox{ or } \hskip 0.5 cm\frac{d}{dt} {\bf J}^T = 0,
\end{equation}
then the electrical or thermal conductivity would be infinite.
Even if the charge and energy currents were not themselves conserved, the electrical or thermal
conductivity would still be infinite, if there were some other conserved quantities
with non-zero `overlap' (in a sense to be made precise in Eq. (\ref{overlap})) with the charge or energy current.
Hence, finite conductivities only occur when the corresponding currents have no overlap
with any conserved quantities.\cite{Mazur1969, Suzuki1971, ZotosNaefPrelovsek1997} 

In addition to total charge and energy there are other globally conserved quantities (we will interchangeable call them charges) for the fixed point action of a hyperconductor in Eq. (\ref{eqn:LL-action}). 
There are 47 conservation laws at the
asymmetric shorter Leech fixed point that 
are important for transport: the charges of the right- and left-movers in
each channel as well as the total energy.\footnote{In fact, there are an infinite number of conserved charges of the Luttinger liquid action describing the hyperconductor fixed point which take the form of products of the chiral current operators defined Eq. (\ref{chiralcurrents}). These additional charges have vanishing overlap with the chiral currents and momentum operator to lowest order in the scattering interaction, and so make subleading contributions to the conductivity and will be ignored.}
We now discuss these conservation laws,
as well as the relaxation mechanisms due to irrelevant perturbations of the fixed point
that are required to make these conductivities finite.

Continuous translation symmetry of the parent non-relativistic theory, whose low-energy effects are captured by $S_{\rm lin}$, gives a globally conserved charge (total momentum), here written in fermionic language:
\begin{align}
    P &= P_0  + P_D, \\
    P_0 &= k_F \sum_I \left (N^R_{I} - N^L_{I} \right ), \\
    P_D &= \int_x \Big[\psi^\dagger_{R, I} (i \partial_x \psi_{R,I}) + \psi^\dagger_{L, I} (i \partial_x \psi_{L, I}) \Big],
\end{align}
where $N^R_{I}$, $N^L_{I}$ are, respectively, the number operators of the right-moving and left-moving Dirac fermions in channel $I$:
\begin{equation}
N^{R,L}_I= \int_x \,\,\psi^\dagger_{R/L,I} \psi^{}_{R/L,I}.
\end{equation}
$P_D$, as suggestively named, is the momentum of a Dirac fermion theory also described by $S_{\rm lin}$, but where $\psi^\dagger_{R,I} \, \left (\psi^\dagger_{L,I}\right )$ creates a right-moving (left-moving) fermion about zero momentum instead of the Fermi point $k_{F,I} \left (-k_{F,I} \right )$.
From the perspective of the low-energy theory, the total momentum operator $P$ arises from two separately conserved emergent symmetries of $S_{\rm lin}$: the first is generated by a chiral rotation of the right- and left-moving fermions by the ``angle" $k_F$ while the second is generated by continuous translations in the linearized Dirac theory.  
$P_0$ accounts for the large momenta $\sim k_F$, while $P_D$ accounts for deviations from the Fermi surface.

These expressions can be rewritten in bosonic form:
\begin{align}
N^{R}_{I} &= \frac{1}{2\pi} \int_x  \partial_x \phi_I, \\
N^{L}_{I} &= \frac{1}{2\pi} \int_x \partial_x \phi_{N+I},
\end{align}
and
\begin{equation}
P_D = \frac{1}{4\pi} \int_x K_{IJ} \partial_x \phi_I \partial_x \phi_J.
\end{equation}
The fermionic and bosonic expressions for $P=P_0 + P_D$ are the integrals
over all space of the component $T^{tx}$ of the energy-momentum tensor derived
via Noether's theorem from, respectively, the fermionic Eq. (\ref{relativistic})
and bosonic Eq. (\ref{eqn:LL-action}) forms of the effective action.

The fixed point action $S_{\rm b}$ has emergent $U(1)^N_L \times U(1)^N_R$ chiral symmetries
($\phi_I \rightarrow \phi_I + c_I$) generated by the charges $Q^{R/L}_{I}$:
\begin{equation}
Q^{R,L}_{I} = e N^{R/L}_I.
\end{equation}
The continuity equation for each chiral charge and the equations of motion for the bosonic fields allow us to obtain the corresponding currents:
\begin{align}
\label{chiralcurrents}
J^{e}_{R,I} &= \frac{e}{2\pi}  V_{IJ} \int_x \partial_x \phi_J,\\
J^{e}_{L,I} &=  -\frac{e}{2\pi}  V_{N+I,J} \int_x \partial_x \phi_J.
\end{align}
The total electrical and thermal currents are then given by:
\begin{align}
    J^{e} &=  \sum_{I=1}^{N} \left(J^{e}_{R,I} +  J^{e}_{L,I}\right), \\
    J^{T} &= - \frac{1}{4\pi} \sum_{I,J,L=1}^{2N} V_{IJ} K_{II} V_{IL} \int_x \partial_x \phi_J\partial_x \phi_L,
\end{align}
where the Hamiltonian,
\begin{align}
H = {1 \over 4 \pi} \int_x V_{IJ} \partial_x \phi_I \partial_x \phi_J,
\end{align}
and corresponding thermal continuity equation gives $J^T$.
We study the case when all of the eigenvalues of $V_{IJ}$ are the same,
so that the Dirac momentum $P_D$ is equal to the thermal current $J^T$.

Particle-hole symmetry breaking band-curvature effects couple the electrical and thermal currents to one another. 
For completeness, we give, in fermionic form,
the corresponding corrections to the expressions for the currents:
\begin{align}
    \delta J^{e} &= g\frac{e}{m} P_D, \\
    \delta J^{T} &= \frac{g}{m} \sum_I \int_x \, \Big[ \left (\partial_t \psi^\dagger_{R,I} \right ) \partial_x \psi_{R,I} + \left ( \partial_x \psi^\dagger_{R,I} \right  ) \partial_t \psi_{R,I} \nonumber \\
    & +  \left (\partial_t \psi^\dagger_{L,I} \right ) \partial_x \psi_{L,I} + \left ( \partial_x \psi^\dagger_{L,I} \right  ) \partial_t \psi_{L,I} \Big].
\end{align}
In an operator formalism, the time derivative of the fermion operator above is computed by taking the commutator of the fermion operator with the Hamiltonian $H$.
If the fermions have quadratic dispersion, so that there are no higher-order corrections
to these expressions for the currents, the action is Galilean-invariant. 
The band curvature corrected electrical current then gives the expected relation between the total electrical current and total momentum, $J^{e} + \delta J^{e} = \frac{e}{m}P$.
Band curvature effects that do not break particle-hole symmetry introduce corrections to
$J^e$ that are odd in the $\phi_I$ and corrections to $J^T$ that are even in the $\phi_I$.
These and other corrections due to band curvature are interesting and deserve further study (see Ref. \onlinecite{ImambekovSchmidtGlazman2012review} for a review), however, we focus upon the linearly dispersing regime in this paper.

To summarize, the fixed point action $S_{\rm b}$ has 47 individually conserved quantities, $Q^{R,L}_{I}$ and
$P_D$, that generally have non-zero overlap with the electrical and thermal currents.
One linear combination of these conserved quantities, the total electrical charge $Q = \sum(Q^{R,}_{I} + Q^{L}_{I})$, will always\footnote{Assuming that the system is not coupled to
an external superconductor to violate charge conservation or driven to violate energy conservation.} remain conserved, but it has no overlap with either the electrical or thermal currents and so it does not prevent their decay.  
The other 46 conservation laws must be broken in order for the system to have finite electrical and thermal
conductivities.

\subsection{Relaxation Mechanisms}

To see the relation between the conductivity and conservation laws, it is
helpful to consider the most general expression for the real part of the
optical conductivity:\cite{SirkerPereiraAffleck2011}
\begin{align}
    \sigma'(\omega,T) = 2\pi D(T) \delta(\omega) + \sigma_{\rm reg}(\omega,T), 
\end{align}
where $D(T)$ is the so-called Drude weight. If $D(T)$ is finite, it signals that the DC conductivity is infinite. Using Mazur's inequality,\cite{Mazur1969, Suzuki1971} Zotos, Naef, and Prelovsek pointed out in Ref. \onlinecite{ZotosNaefPrelovsek1997} the following
implication of conserved charges for electrical charge transport:
\begin{align}
\label{mazur}
    D(T) & \geq \frac{1}{2 L T} \frac{\sum_k \langle J^{e} Q_k \rangle^2}{ \langle Q_k^2 \rangle },
\end{align}
where $L$ is the length of the system.
The angled brackets denote the thermodynamic average and the right-hand side of Eq. (\ref{mazur}) is independent of time because the $Q_k$ are conserved quantities.
This inequality says that in the presence of conserved charges $Q_k$ which have non-zero overlap with $J^{e}$, the electrical current does not completely relax, and the system has dissipationless charge flow even at finite temperature $T$.
(See Eq. (\ref{overlap}) for an equivalent notion of an `overlap' which is the one that we adopt in this paper.)
A similar inequality and conclusion applies for the thermal current $J^T$.

It follows that to fully relax the electrical and thermal currents a system must break all conservation laws, apart from the conservation of total charge and total energy,
which have vanishing overlap with the electrical and thermal currents.
At zero-temperature and zero
frequency, the fixed point theory $S_{\rm b}$ determines the response of the system.
Since this theory has the 47 conservation laws described above,
it has infinite conductivity. Note that, in a time-reversal invariant 23-channel
Luttinger liquid, we would only need to break 24 conservation laws
since the time-reversal symmetric conserved quantities would ordinarily have vanishing overlap
with the electrical current; but the asymmetric Leech liquid hyperconductor is not time-reversal invariant.

At finite temperature and frequency, irrelevant perturbations can have an effect on the
response functions of the system. The bulk of this paper is a discussion of the effects of such perturbations. In particular, we answer two questions: Which operators can relax the currents? Which are the most important ones?

In order to break the conservation of the Dirac momentum $P_D$ and the chiral electrical currents $\lbrace J^{e}_{R/L,I} \rbrace$,
we need to include physical processes that (1) break continuous translation symmetry with respect to the low-energy effective theory $S_{\rm b}$ and (2) break particle number conservation within each channel, but (3) conserve total charge and energy.
Umklapp scattering at incommensurate fillings and disorder break continuous momentum conservation and generally break
the conservation of the chiral currents in individual channels, and so we focus on
them here.

Umklapp processes scatter some number of right-movers into left-movers so that the total momentum change is a reciprocal lattice vector.
The most general umklapp term is specified by a vector of integers $ m^{(\alpha)}_I, I=1, \ldots, 2N$:
\begin{align}
\label{umklappscattering}
    H_{\rm u} =& \sum_{\alpha} H^{\rm u}_{\alpha} \cr
    = & \sum_{\alpha} \Big[ h^{\rm u}_{\alpha} + {\rm h.c.} \Big] \cr
    =&  - \sum_\alpha \lambda_\alpha \int_{x} \left[ {1 \over a^2} e^{i m^{(\alpha)}_I k_{F,I} x - i p^{(\alpha)} G x} e^{ i m_J^{(\alpha)} \phi_J} + {\rm h.c.} \right],\cr
\end{align}
where $\lambda_\alpha$ is the coupling constant, $G$ is a basis vector of the reciprocal lattice, $a$ is a short-distance cutoff,\footnote{Typically, the presence of a multiplicative prefactor proportional to a power of the short-distance cutoff $a$ is understood when writing vertex operators of the form, $\exp\Big(i m_J^{(\alpha)}\Big)$. We retain it here when writing vertex operators of scaling dimension equal to 2 to avoid confusion. See Ref. [\onlinecite{ginspargappliedcft1991}] for further details.} and the Einstein summation convention is employed. 
Here, the operator $X$ to which we referred in our general remarks in Sec. \ref{sec:remarks} is
$X=e^{ i m_J^{(\alpha)} \phi_J}$.
The most important umklapp processes at low energies are those for which the corresponding operators
$X=e^{ i m_J^{(\alpha)} \phi_J}$ have the lowest scaling dimension.
In the asymmetric shorter Leech hyperconductor studied in this paper, such operators have scaling dimension $(\Delta_R, \Delta_L) = (3/2, 1/2)$, so they are marginal.
The integer $p^{(\alpha)}$ is the ``order" of the umklapp process, or the number of Brillouin zone foldings after which the momentum $m^{(\alpha)}_I k_{F,I}$ is again in the first Brillouin zone. Thus, $p^{(\alpha)}$ is actually fixed by $m^{(\alpha)}_I k_{F,I}$, but we will retain it as a formally free parameter. At commensurate filling, there is always a $p^{(\alpha)}$ such that $m^{(\alpha)}_I k_{F,I} = p^{(\alpha)} G$, but we work more generally.
Without loss of generality, we may take the difference $m_I^{(\alpha)} k_{F,I} - p^{(\alpha)} G \in [0, 2 \pi)$ where the lattice constant has been set to unity. 
Charge conservation is maintained by requiring equal numbers of creation and annihilation
operators: $ \sum_{I=1}^{N} m^{(\alpha)}_I = \sum_{I=1}^{N} m^{(\alpha)}_{N+I}$.

While any single umklapp process $H^{\rm u}_{\alpha}$ might break the conservation of individual currents (e.g., $[H^{\rm u}_{\alpha},J^e_{R/L, I}] \neq 0$), a linear combination of currents might still be conserved.\cite{Shimshoni2003}
(The linear combination corresponding to total charge is always conserved, however, it has no overlap with the total electrical current.). 
That is why our model generally requires at least 46 carefully chosen umklapp processes, i.e., $m^{(\alpha)}_I$ vectors to break all conservation laws. 
Such a requirement is not unreasonable.
In the spirit of effective field theory, we expect all operators consistent with symmetry to be present in the low-energy effective action.
We simply focus on the minimal set of scattering processes that dominate the low-energy physics. 
See the accompanying Mathematica file for explicit expressions of the $m_I^{(\alpha)}$ that we choose to study.

To study whether some linear combination (other than the total charge) $a_I J_I$ with $J_I = J^e_{R,I}$ for $I = 1, \ldots N$ and $J^e_I = J^e_{L,I - N}$ for $I = N+1, \ldots, 2N$ is also conserved, we compute the equal-time commutators:
\begin{align}
 \label{generalcurrentcommutator}
 [H^{\rm u}_{\alpha}, a_I J^e_I] = & i a_I b_I^{\alpha} h^{\rm u}_{\alpha} + {\rm h.c.},
 \end{align}
 where the vectors $b_I^{\alpha}$ are defined by,
 \begin{align}
 b_I^{\alpha} = \Big(e \lambda_\alpha {\rm sgn}(N-I) {\rm sgn}(N-J) V_{IJ}\Big) m_J^{(\alpha)},
 \end{align}
 and we define ${\rm sgn}(X) = +1$ for $X \geq 0$ and ${\rm sgn}(X) = - 1$ for $X < 0$.
We ask whether there exist solutions $a_I = \vec{a} \in \bR^{2N} - \{ \mathbf{0}\}$, such that $\forall \alpha,  a_I b^{\alpha}_I = 0$.
All umklapp operators preserve total $U(1)$ electrical charge, therefore the vectors $m^{(\alpha)}_I$ specifying them can span at most a $2N-1$ dimensional space. 
The linear equations, $a_I b_I^\alpha = 0$, say that $\vec{a}$ is orthogonal to this space. 
It follows that when the number of linearly independent umklapp terms $N_U$ ($\alpha=1, \ldots, N_U$) equals $2N-1$, $\vec{a}$ lies in the 1-dimensional space corresponding to total charge, and so no non-trivial conserved linear combination of the currents exists. 

Disorder can also relax the electrical and thermal currents by violating
conservation laws.
A generic disorder-mediated backscattering term takes the form:
\begin{align}
\label{disorderscattering}
    H_{\rm dis} =  & \sum_\alpha \lambda^{\rm dis}_{\alpha} H^{\rm dis}_{\alpha} \cr
    = & \sum_{\alpha} \lambda^{\rm dis}_\alpha \int_x \Big[{\xi_\alpha}(x) {1 \over a^2} e^{i m^{(\alpha)}_I \phi_I} + {\rm h.c.}\Big],
\end{align}
where $\alpha$ indexes the various backscattering terms specified by $m^{(\alpha)}_I \in \bZ$. 
At low temperatures, the most important backscattering processes are again
due to the dimension $(\frac{3}{2},\frac{1}{2})$ operators $e^{i m^{(\alpha)}_I \phi_I}$ introduced in Eq. (\ref{umklappscattering}).
However, due to randomness in ${\xi_\alpha}(x)$, their effect is weaker than that of
uniform umklapp terms. (In the general remarks in Sec. \ref{sec:remarks}, the operator $X=e^{i m^{(\alpha)}_I \phi_I}$
in Eq. (\ref{disorderscattering}).)

For simplicity, we will take all the couplings $\lambda^{\rm dis}_\alpha =  \lambda^{\rm dis}$ equal and $\overline{\xi_\alpha(x) \xi^*_\beta(x')} = \delta_{\alpha\beta} D \delta(x-x')$ with $\overline{\xi_\alpha}(x) = 0$, where the overline denotes disorder averaging.
Then, we use the replica trick to integrate out the disorder, thereby obtaining the
following term in the replicated action:
\begin{align}
    S_{\rm dis-avg} = (\lambda^{\rm dis})^2 D \sum_{A,B} \sum_{\alpha} \int_{t,t'} \int_x
{1 \over a^4} e^{i m^{(\alpha)}_I ({\phi^A_I}(t) - {\phi^B_I}(t'))}. 
\end{align}
For a dimension $(\frac{3}{2},\frac{1}{2})$ operator $e^{i m^{(\alpha)}_I \phi_I}$,
the coupling $(\lambda^{\rm dis})^2 D$ of the interaction in the replicated theory has scaling dimension equal to $-1$. 
Hence, the interaction is irrelevant and its effects are
formally subleading compared to the uniform umklapp terms considered above.
However, in the commensurate case, umklapp terms commute with $P_D$; disorder is
the leading effect that violates conservation of $P_D$, thereby leading to finite
thermal conductivity. Meanwhile, in the incommensurate case, the effects of
uniform umklapp terms are exponentially-suppressed at low temperatures,
and disorder becomes the leading effect that relaxes both electrical and thermal
currents at low temperatures.

In summary: for a pure system at commensurate filling, the Dirac momentum $P_D$ is not relaxed, however, there is no overlap between the chiral electrical currents $J^e_I$ and $P_D$ when particle-hole symmetry is preserved. 
Thus, we need $45$ umklapp operators to relax the electrical current.
When particle-hole symmetry is broken by band-curvature corrections at commensurate filling, $\langle J^e P_D \rangle \neq 0$, so both the electrical and thermal conductivities diverge.
When the filling is incommensurate or disorder is present, particle-hole symmetry is broken, so there is generally an overlap between the electrical currents and the Dirac momentum.
However, $P_D$ does not generally commute with an umklapp process at incommensurate filling or a disorder-mediated scattering interaction, thereby allowing momentum relaxation. 
In this case, both the electrical and thermal transport coefficients can be finite in the presence of $46$ scattering interactions. The additional interaction arises from the additional conserved charge $P_D$. To see this one must generalize the previous argument by writing the commutator in Eq. (\ref{generalcurrentcommutator}) as a total derivative.

\subsection{Memory Matrix}
\label{memorymatrixsection}

The details of the memory matrix formalism can be found in Refs. [\onlinecite{Forster1975, Giamarchibook, Shimshoni2003, Hartnolllectures2013, LucasSachdev2015MM}]; we merely observe that it is well-suited for computing transport coefficients in the hydrodynamic regime: when there are globally conserved quantities (energy, electrical charge) that propagate diffusively. Unlike a direct application of the Kubo formulae it makes the role of these conservation laws transparent. 
In essence, it is a reorganization of the perturbative expansion of the current-current correlation functions of interest.\cite{SirkerPereiraAffleck2011}

We choose as a complete basis of conserved quantities the set $\lbrace {\cal Q}_p \rbrace = \lbrace J^e_{R, 1},...J^e_{R, N}, J^e_{L, 1},...J^e_{L, N-1}, P_D \rbrace$. $J^e_{L,N}$ can be excluded because total charge is
always conserved, so a correlation function involving $J^L_N$ can be obtained from an expression involving the other currents.
There is a notion of a symmetric inner product on the vector space of conserved quantities provided by the static susceptibility matrix:
\begin{align}
\label{overlap}
\hat{\chi}_{pq} = & \left ( {\cal Q}_p | {\cal Q}_q \right ) \cr
            \equiv & \frac{1}{L} G^R_{{\cal Q}_p {\cal Q}_q}(\omega=0).
\end{align} 
The retarded Green's functions $G^R_{{\cal Q}_p {\cal Q}_q}(\omega)$ are calculated at temperature $T$ (left implicit in the definitions below) and evaluated at real frequency $\omega$.
(Recall that there is no momentum dependence in the static susceptibility matrix $\hat{\chi}_{pq}$ because the conserved charges are obtained by integrating densities over all space.)
Thus, the static susceptibility may be used to define the notion of an `overlap' between two conserved quantities. 
Note that the real-time thermodynamic correlation functions involved in Mazur's inequality Eq. (\ref{mazur}) are non-zero if and only if the corresponding static susceptibilities are non-zero.

The memory matrix $\hat{M}(\omega)$ has contributions from each separate umklapp and disorder-mediated scattering process, both labeled by $\alpha$. 
We schematically write this as:
\begin{align}
\hat{M}(\omega) =& \sum_\alpha \Big(\lambda_\alpha^2 \hat{\cal M}^{\rm u}_\alpha(\omega) + (\lambda^{\rm dis}_\alpha)^2 D \hat{\cal{M}}^{\rm dis}_\alpha(\omega)\Big), \\ 
(\hat{\cal M}^{\rm u})^{pq}_\alpha =& \frac{1}{L}\frac{\langle F^{\rm u}_{p, \alpha}; F^{\rm u}_{q, \alpha} \rangle_\omega - \langle F^{\rm u}_{p, \alpha}; F^{\rm u}_{q, \alpha} \rangle_{\omega=0} }{ i \omega}, \\
(\hat{\cal M}^{\rm dis})^{pq}_\alpha =& \frac{1}{L}\frac{\langle F^{\rm dis}_{p, \alpha}; F^{\rm dis}_{q, \alpha} \rangle_\omega - \langle F^{\rm dis}_{p, \alpha}; F^{\rm dis}_{q, \alpha} \rangle_{\omega=0} }{ i \omega}.
\end{align}
Here, $F^{\rm u}_{q, \alpha} = {i \over \lambda_\alpha} \left[H^{\rm u}_{\alpha}, {\cal Q}_q \right]$, $F^{\rm dis}_{q, \alpha} = {i \over \lambda^{\rm dis}_\alpha \sqrt{D}} \left[H^{\rm dis}_{\alpha}, {\cal Q}_q \right]$, and ${\cal Q}_q$ is a conserved charge (either $J^e_{R/L, I}$ or $P_D$). 
$\langle F^{\rm u}_{p, \alpha}; F^{\rm u}_{q, \alpha} \rangle_\omega$ and $\langle F^{\rm dis}_{p, \alpha}; F^{\rm dis}_{q, \alpha} \rangle_\omega$ are retarded finite-temperature Green's functions evaluated to leading order using $S_{\rm b}$ in Eq. (\ref{eqn:LL-action}).
$\lambda_\alpha$ and $\lambda^{\rm dis}_\alpha$ parameterize the umklapp scattering and coupling to disorder, respectively, and $D$ is the disorder variance of Gaussian-correlated disorder.
As mentioned above, we take $\lambda_\alpha = \lambda$ and $\lambda_\alpha^{\rm dis} = \lambda^{\rm dis}$ for all $\alpha$ for simplicity.
$\hat{{\cal M}}^{\rm u}$ contains the contributions to the memory matrix from umklapp scattering, while $\hat{{\cal M}}^{\rm dis}$ contains the contributions from the disorder-mediated interaction.
We stress that the form of the memory matrix given above is correct to leading order in the scattering interaction.
See Refs. [\onlinecite{Forster1975, Giamarchibook, Shimshoni2003, Hartnolllectures2013, LucasSachdev2015MM}] for further discussion.

The label $\alpha$ also specifies the momentum mismatch of an incommensurate scattering process,
\begin{align}
\Delta k_{\alpha} \equiv m^{(\alpha)}_I k_{F,I} - p^{(\alpha)} G \in [0, 2\pi),
\end{align}
for unit lattice constant,
and the vector of integers $m^{(\alpha)}_I$ that defines the umklapp process.
The vectors $m^{(\alpha)}_I$, in turn, help determine, along with the matrix $V_{IJ}$, the right and left scaling dimensions $(\Delta_R, \Delta_L)$ of the operators entering scattering interactions in Eqs. (\ref{umklappscattering}) and (\ref{disorderscattering}).
Recall that we choose to take the Fermi vectors in all channels to be equal, $k_{F,I} = k_F$.

The conductivities associated to the various charges ${\cal Q}_p$ are encoded in the matrix,
\begin{equation}
    \label{conductivitymatrix}
    \hat{\sigma}(\omega) = \hat{\chi} \left( \hat{N}+ \hat{M}(\omega) - i \omega \hat{\chi} \right)^{-1} \hat{\chi},
\end{equation}
where 
\begin{align}
(\hat{N})_{pq} \equiv \hat{\chi}_{p \dot{q}} = \Big({\cal Q}_p, i [\sum_\alpha (H^{\rm u}_{\alpha} + H^{\rm dis}_\alpha), {\cal Q}_q])\Big).
\end{align}
We show in Appendix \ref{Nmatrixappendix} that, at least to quadratic order in the umklapp $\lambda$ and disorder $\lambda^{\rm dis}$ couplings, $\hat{N} = 0$.

The electrical conductivity $\sigma$ is determined by the $(2N - 1) \times (2N -1)$ submatrix $\hat{\sigma}_{J_I^e, J_J^e}$.
The thermoelectric conductivity $\tilde{\alpha}$ is determined by the $(2N - 1)$-dimensional vector $\hat{\sigma}_{J_I^e, P_D}/T$.
The thermal conductivity $\kappa = {\hat{\sigma}_{P_D, P_D} \over T} - {\tilde{\alpha}^2 T \over \sigma}$. 
For commensurate fillings and in the disorder-dominated regime, the thermoelectric conductivity can be ignored to leading order so that the thermal conductivity is equal to the $P_D - P_D$ component of $\hat{\sigma}$.

\section{Hyperconductor Transport}
\label{PMtransport}

We now assemble the conductivity matrix $\hat{\sigma}$. 
The first ingredient is the static susceptibility matrix, which takes the following form:
\begin{align}
\label{staticelectricmatrix}
\hat{\chi}_{J^e_I J^e_J} = & {e^2 \over 4 \pi} {\rm sgn}(N - I) {\rm sgn}(N - J) V_{IJ}, \\
\label{staticelectricmomentummatrix}
\hat{\chi}_{J^e_I P_D} =  & 0, \\
\label{staticmomentummatrix}
\hat{\chi}_{P_D P_D} = & {N \pi^2 T^2 \over 6},
\end{align}
where there is no sum over $I$ and $J$ and we have computed to zeroth order in any perturbation to $S_{\rm b}$.
See Appendix \ref{staticappendix} for details on the calculation of the static susceptibilty matrix and the auxiliary Mathematica file for the explicit expression for $V_{IJ}$.
See Appendix \ref{memorymatrixcomputations} for details on the evaluation of the memory matrix elements.

In the following two sections, we study the contributions to the conductivity in systems at commensurate and incommensurate fillings in the presence of both umklapp scattering and disorder.
For the most part, we focus upon the decoupled surface subspace within the hyperconductor phase, however, we provide the more general expressions for the DC conductivities where appropriate. 

\subsection{Commensurate Fillings}

If the electron filling is commensurate with the lattice, $k_F$ divided by the reciprocal lattice basis vector is a rational fraction, and so the momentum mismatch $\Delta k_\alpha$ in any umklapp scattering process may vanish.
Umklapp scattering interactions with $\Delta k_\alpha = 0$ provide the dominant contribution to the electrical conductivity matrix.
Thus, we consider $S_{\rm b}$ together with $45$ umklapp terms, all with $\Delta k^{(\alpha)}_p = 0$. As argued earlier, the most important umklapps are those with total scaling dimension $(\Delta_R, \Delta_L) = (3/2,1/2)$.

\subsubsection{DC Conductivity}

We first note that $F^{\rm u}_{P_D, \alpha}$ vanishes when $\Delta k^{(\alpha)}=0$ ,
along with all the memory matrix elements involving it. 
This tells us that the dynamics of the electrical current-carrying excitations decouple from the thermal carriers (with $P_D$ remaining conserved) at commensurate fillings without disorder.
In computing the electrical conductivity, it is sufficient to choose $\lbrace J^e_I \rbrace $ as the complete basis of hydrodynamic modes.
The conservation of $P_D$ in the linearly-dispersing regime also implies that the thermal conductivity $\kappa$ is infinite in a pure system since $\left( P_D | J^{T} \right) \neq 0$. 
At commensurate fillings, disorder is the leading effect that
causes finite thermal conductivity, as we discuss.

To obtain the DC conductivity at commensurate fillings, we need the memory matrix elements obtained in Appendix \ref{commensurateumklapp}:
\begin{align}
(\hat{\cal M}^{\rm u})^{J^e_I J^e_J}_\alpha(T) = {\pi^4 \over 32} U_{J^e_I, \alpha} U_{J^e_J, \alpha} T,
\end{align}
where the finite, non-zero coefficients, $U_{J^e_I, \alpha} U_{J^e_J, \alpha} \propto e^2$ are defined in Eq. (\ref{umklappcoefficients}).
This immediately gives the DC electrical conductivity,
\begin{align}
\label{electricDCcommensurate}
\sigma(T) \propto {e^2 \over \lambda^2} {1 \over T}.
\end{align} 
As promised, the electrical resistivity vanishes linearly in temperature.
Note that the dimensionless proportionality constants in Eq. (\ref{electricDCcommensurate}) and in subsequent conductivity formulas are finite and non-zero.\footnote{In general, inversion of the $46 \times 46$ memory matrix is computational difficult and so a precise determination of the overall numerical constant prefactors is currently out of reach. Nevertheless, we have checked that the memory matrix is generically non-singular and so we may safely understand the contributions to the relaxation of the various currents by scaling out any dimensionful quantities from the memory matrix. The remaining numerical matrix then merely contributes a finite constant whose overall value we do not determine.}

We have neglected band curvature terms in the preceding and subsequent calculations by
working with the linearized action in Eq. (\ref{eqn:LL-action}). 
Their inclusion does not lead to
finite thermal conductivity since any non-oscillatory term will commute with $P_D$.
However, particle-hole symmetry-breaking band curvature terms will mix $P_D$ and $J^e_I$, thereby
leading to infinite electrical conductivity so long as $P_D$ is conserved.

Disorder, on the other hand, does cause $P_D$ to decay. 
While it gives a subleading contribution to the electrical conductivity in the commensurate case -- disorder contributes the ${\cal O}(T^2)$ correction in Eq. (\ref{disorderelectricmemory}) to the DC electrical memory matrix elements -- it is the leading contribution to
the relaxation rate of the thermal conductivity:
\begin{equation}
    \label{disordercommensurate}
    \kappa(T) \propto \left(\frac{1}{D (\lambda^{\rm dis})^2}\right) \frac{1}{T},
\end{equation}
where we have used the static susceptibility matrix in Eq. (\ref{staticmomentummatrix}), the disorder memory matrix elements in Eq. (\ref{disordermomentummemory}), and the fact that $\kappa T$ is equal to the ${P_D}-{P_D}$ component of the conductivity tensor $\hat{\sigma}$ when the thermoelectric coefficient vanishes (to leading order).

Eqs. (\ref{electricDCcommensurate}) and (\ref{disordercommensurate}) constitute a violation of the Wiedemann-Franz ``law."
Marginal umklapp scattering is the leading low-temperature relaxation mechanism for the electrical current, while ${\cal O}(1)$ irrelevant disorder is the leading relaxation mechanism for the thermal current at commensurate fillings. 
In this case, the Lorentz ratio,
\begin{align}
{\cal L} = {\kappa \over \sigma T} \propto {\lambda^2 \over e^2 D (\lambda^{\rm dis})^2} {1 \over T}
\end{align}
diverges as $T \rightarrow 0$.

Remaining within the hyperconductor phase, but departing from the decoupled surface, the exponents for the electrical and thermal conductivities will generally be modified to the form: $\sigma \propto 1/T^{1 - 2(2-\Delta_X)}$ and
$\kappa \propto 1/T^{1 - 2(2-\Delta_X)}$, where deviations of $\Delta_X$ from $2$ encode the shift of the scaling dimensions of the scattering processes away from marginality.

\subsubsection{AC Conductivity}

The AC conductivities at commensurate fillings are found similarly.
From Appendix \ref{commensurateumklapp},
\begin{align}
\label{commensurateAC}
(\hat{\cal M}^{\rm u})^{J^e_I J^e_J}_\alpha(\omega) = U_{J^e_I, \alpha} U_{J_J^e, \alpha} \Big[ {\pi^2 \over 32} \omega + i {\pi \over 16} \omega \log(a_2 \omega) \Big],
\end{align}
where $a_2$ is proportional to the short-distance cutoff $a$.
Therefore, the AC electrical conductivity at $T \ll \omega$ takes the form:
\begin{align}
\label{ACcommensurate}
\sigma(\omega) \propto {e^2 \over i \omega\Big(c_1 + c_2 \log(a_2 \omega)\Big) + c_3 \omega}, 
\end{align}
for constants $c_1, c_2$ and $c_3$.
The finite contribution to the real part of the electrical AC resistivity has given the Drude peak finite width.

Disorder is required for finite AC thermal conductivity.
Using the memory matrix element in Eq. (\ref{disordermomentummemory}), we find:
\begin{align}
\kappa(T/\omega \ll 1) \propto {T^3 \over i c_4 \omega T^2 + c_5 D \omega^4}, 
\end{align}
for constants $c_4$ and $c_5$.

\subsection{Incommensurate Fillings}

When the filling is incommensurate, there is no scattering process for which $\Delta k_\alpha = 0$.
In this case both the electrical and thermal conductivities are generally finite and so we use the charge basis $\lbrace {\cal Q}_p \rbrace = \lbrace J^e_{R, 1},...J^e_{R, N}, J^e_{L, 1},...J^e_{L, N-1}, P_D \rbrace$.
Band-curvature corrections contribute subleading terms to the temperature dependence and will not be considered. 

The $\Delta k_\alpha$ associated to the $46$ umklapp scattering processes defined by the $m_I^{(\alpha)}$ vectors are all generally different from one another.
Nevertheless, we set $\Delta k_\alpha = \Delta k$ for all $\alpha$ in the presentation of the results below.

\subsubsection{DC Conductivity}
\label{incommensurateDCsection}

The memory matrix elements for umklapp scattering at incommensurate filling is provided in Appendix \ref{incommensurateumklapp} whose results we quote below.

Infinitesimally close to commensurate filling, $\omega \leq \Delta k \ll T$, we may borrow our previous results computed precisely at commensurate filling with the understanding that $\Delta k \neq 0$ in the expression for $F^{\rm u}_{P_D, \alpha}$ in Eq. (\ref{umklappcomms2}).
The leading contribution to the electrical conductivity is unchanged from Eq. (\ref{electricDCcommensurate}). 
However, the thermal conductivity is now finite even in the absence of disorder,
\begin{align}
\kappa(T) \propto {T^2 \over \lambda^2 \Delta k^2}.
\end{align}
As expected, the thermal conductivity is divergent as commensurability is restored, $\Delta k \rightarrow 0$.
The Lorentz ratio is a decreasing function of $T^2$ in the regime $\Delta k \ll T$ as the temperature is decreased.

As the temperature is lowered, we enter the regime $T \ll \Delta k$ in which the DC electrical and thermal memory matrix elements take the asymptotic low-temperature form:
\begin{align}
\label{incommen}
(\hat{\cal M}^{\rm u})^{pq}_\alpha(T) = & {\pi^2 \over 32} U_{p, \alpha} U_{q, \alpha} {\Delta k^2 \over T} e^{- {\Delta k \over 2 T}}.
\end{align}
The resulting DC electrical and thermal conductivities for $T \ll \Delta k$:
\begin{align}
\sigma(T) \propto & {e^2 \over \lambda^2} {T \over \Delta k^2} e^{{\Delta k \over 2 T}}, \cr
\kappa(T) \propto & {1 \over \lambda^2} {T^4 \over \Delta k^4} e^{{\Delta k \over 2 T}}.
\end{align}
In this case, the Lorentz ratio,
\begin{align}
{\cal L} \propto {T^2 \over e^2 \Delta k^2},
\end{align}
vanishes as $T\rightarrow 0$ in the absence of disorder.
If we had considered instead a more generic model in which the Fermi momenta were not identical, the $\Delta k$ would then no longer be same.
This would imply that the leading contribution to the memory matrix in Eq. (\ref{incommen}) would be dominated by the contribution with minimal $\Delta k$.

Disorder, if present, eventually dominates the low-temperature transport.
The disorder DC electrical and thermal memory matrix elements derived in Appendix \ref{disordermemory}: 
\begin{align}
\label{disorderelectricmemory}
(\hat{\cal M}^{\rm dis})^{J^e_{I} J^e_{I}}_\alpha = & {2 \pi^3 \over 3} \tilde{U}_{J^e_I, \alpha} \tilde{U}_{J^e_J, \alpha} T^2,\\
\label{disorderelectricmomentummemory}
(\hat{\cal M}^{\rm dis})^{J^e_{I} P_D}_\alpha = & 0, \\
\label{disordermomentummemory}
(\hat{\cal M}^{\rm dis})^{P_D P_D}_\alpha = & {8 \pi^5 \over 5} \tilde{U}_{P_D, \alpha} \tilde{U}_{P_D, \alpha}T^4,
\end{align}
where the coefficients $\tilde{U}_{p, \alpha} \tilde{U}_{q, \alpha}$ are defined in Eqs. (\ref{disordercoefficients}).
For generic, perturbative values of the couplings, the disorder-dominated regime occurs when the exponentially-vanishing contribution to the memory matrix in Eq. (\ref{incommen}) is overcome by the disorder-dominated contribution above.
The resulting electrical and thermal conductivities in the presence of disorder for temperatures $T \ll \Delta k$:
\begin{align}
\sigma(T) \propto & {e^2 \over D (\lambda^{\rm dis})^2} {1 \over T^2}, \cr
\kappa(T) \propto & {1 \over D (\lambda^{\rm dis})^2} {1 \over T}.
\end{align}
Away from the decoupled surface, the low-temperature results will be modified as follows: $\sigma = \kappa/T \propto 1/T^{2 - 2(2-\Delta_X)}$.
In this regime, the Lorentz ratio,
\begin{align}
{\cal L} \propto {1 \over e^2},
\end{align}
is constant, although the gapless metallic phase is certainly {\it not} a Fermi liquid.
The Wiedemann-Franz law is satisfied at the lowest of temperatures for incommensurate fillings because disorder is the dominant relaxation mechanism at incommensurate fillings for {\it both} the electrical and thermal currents. 

\subsubsection{AC Conductivity}

The AC conductivity at incommensurate filling follows straightforwardly from the previous analysis.
For $T \leq \Delta k \ll \omega$, the AC electrical conductivity is unchanged from the previous result in Eq. (\ref{ACcommensurate}).
In fact, the real part of the AC electrical resistivities can be found from inversion of the DC electrical conductivities in Sec. \ref{incommensurateDCsection} by the replacement $T \rightarrow \omega$ in all algebraic prefactors and so we shall not write them out explicitly.

Let us now concentrate on the real part of the AC thermal conductivities.
For $T \ll \Delta k \ll \omega$, 
\begin{align}
\kappa(\omega) \propto {1 \over \lambda^2} {T^3 \over \Delta k^2 \omega}.
\end{align}
For $T < \omega \ll \Delta k$ with $T \ll (\Delta k^2/\omega) \exp({\omega - \Delta k \over 2T})$ and in the absence of disorder the thermal conductivity is dominated by incommensurate umklapp scattering,
\begin{align}
\kappa(\omega) \propto {1 \over \lambda^2} {T^3 \omega \over \Delta k^4} e^{{\Delta k - \omega \over 2 T}},
\end{align}
where we used Eq. (\ref{incommensuratefrequencymemorymatrix}).
Notice the divergent thermal conductivity as $T \rightarrow 0$.
Finally, in the disorder-dominated regime with $T^2 \ll D \omega^3$,
\begin{align}
\kappa(\omega) \propto {1 \over D} {T^3 \over \omega^4}.
\end{align}

\section{Conclusions}
\label{Conclusion}

In this paper, we have determined the DC and AC electrical and thermal conductivity of the one-dimensional hyperconductor phase introduced in Ref. [\onlinecite{Plamadeala2014}] in the presence of umklapp and disorder-mediated scattering.
For instance, we have shown that this metallic phase exhibits a DC conductivity $\sigma \sim 1/T^{1-2(2-\Delta_X))}$ down to $T=0$ without fine-tuning at commensurate fillings, thereby manifesting the non-Fermi liquid nature of the phase.
In addition, we have discussed the relation between conservation laws and transport which has allowed us to provide examples of violations of the Wiedemann-Franz law.
As a simple example, the thermal conductivity is only finite in the presence of disorder, while the electrical conductivity can be finite in a pure system at commensurate filling with only umklapp scattering.
More generally, we have seen how differing relaxation mechanisms for the electrical and thermal currents can result in violations of the Wiedemann-Franz law.

The power-law $\sigma \sim 1/T$ obtains along the `decoupled surface' of the hyperconductor when the interactions determined by $\tilde{V}_{IJ}$ -- see Sec. \ref{PMreview} -- are block diagonal at commensurate fillings. On this surface,
${\Delta_X}=2$. The hyperconductor phase survives within a finite window off the decoupled surface by the addition of off-diagonal terms to $\tilde{V}_{IJ}$ mixing right-moving and left-moving hyperconductor excitations.
Departing from the decoupled surface, but remaining within the hyperconductor phase,
the relaxation of the current is controlled by 46 umklapp scattering operators with conformal dimensions
$\bigl(\frac{3}{2}+\delta,\frac{1}{2}+\delta\bigr)$
so that ${\Delta_X} = 2 + 2\delta$, with $\delta$ determined by the distance from the decoupled surface.
The conductivity will generally behave
$\sigma \sim 1/T^{1-2(2 - {\Delta_X})}$ with ${\Delta_X} > 2$ down to $T=0$. For ${\Delta_X} < 2$, the zero-temperature
perfect metal fixed point is unstable. However, the relevant perturbations are chiral and, therefore, cannot open a gap.
At low temperatures, they may strongly renormalize the velocities, shift the Fermi momenta, or otherwise modify the ground
state (without opening a gap) in such a manner that the dangerous processes can no longer occur. In the marginal case,
$\Delta_X = 2$, such an instability presumably occurs at sufficiently large marginal coupling.

The large marginal coupling limit of this hyperconductor regime is an interesting testing ground for Hartnoll's recently
conjectured\cite{Hartnolldiffusionbound} lower bound
on the diffusion constant, $D \geq \hbar v_F^2/(k_B T)$. This bound applies to systems in the ``incoherent" metallic regime
where there is no overlap between the electrical current and momentum operator.
If satisfied, this lower bound implies an upper bound on the coefficient of the linear in temperature
DC electrical resistivity that we found at commensurate fillings.

The distinction between a hyperconductor and a superconductor is that a hyperconductor
does not have long-ranged order.\footnote{In the case of a 1D system, long-ranged order
is impossible. However, a 1D superconductor develops long-ranged order when in contact
with a 3D superconductor, while a 1D hyperconductor does not. It resists the development of
a proximity effect due to weak coupling to a 3D superconductor.} 
This distinction is not apparent in zero-temperature electrical transport, which is infinite in both cases.
(It does manifest itself in the differential tunneling conductance, which vanishes
algebraically with voltage in the hyperconductor but is strongly suppressed at voltages
below the energy gap in a superconductor -- it would be zero but for Andreev reflection.)
However, the difference between a hyperconductor and a superconductor is clearer in
low-temperature transport. In a superconductor, the electrical resistivity vanishes for
all temperatures below the critical temperature, but in a hyperconductor, the resistivity
increases smoothly, with the temperature dependence described above. In the
incommensurate case, the resitivity is exponentially-small in temperature over
a wide range of temperatures, has the feature of very small (albeit not vanishing)
resistivity without the threat of a sudden large jump at a critical temperature.
While a superconductor conducts electrical current without dissipation even in the presence of impurities
for $T<T_c$, a hyperconductor has non-zero resistivity for $T>0$, but strongly suppressed -- in the
hyperconductor studied here, the impurity contribution is suppressed by a factor $(T/{T_F})^{2\Delta_X-2}$
with $\Delta_X \geq 2$.
Meanwhile, a hyperconductor has radically different thermal transport than
a superconductor. In a superconductor, thermal currents are only carried by excited quasiparticles and
phonons. Therefore, the thermal conductivity divided by the temperature vanishes
with decreasing temperature. In particular, the electronic
contribution to the thermal conductivity of an $s$-wave superconductor has activated form.
In a hyperconductor, on the other hand, the thermal conductivity diverges as a power-law
at the lowest temperatures and diverges exponentially with inverse temperature over
a wide range of temperatures. Thus, the hyperconductor phase, though neither a
superconductor nor a superfluid, has an electrical conductivity that approaches
that of the former and a thermal conductivity that approaches that of the latter.

In the future, we plan to understand the 2D metallic phase that emerges from an array of hyperconductor wires.
This wire array should exhibit diffusive finite-temperature
transport both along and transverse to the wires and be stable to weak disorder.
This paper makes clear the reason why finite conductivities obtain along the wires.
To understand the latter two statements, we need only observe that such an array forms a sort of `chiral transverse Fermi liquid' in the sense that only half of the Fermi surface excitations can hop between wires at the lowest of energies, reminiscent of the chiral metals studied in Refs. [\onlinecite{ChalkerDohmen1995, Balents96, Balents97}] (see Ref. [\onlinecite{SurLee2014}] for related work).
In these works \cite{ChalkerDohmen1995, Balents96, Balents97}, it was found that a collection of wires, each hosting a chiral Fermi liquid (obtained from the edge excitations of a collection of integer quantum Hall systems layered in a transverse direction), exhibits diffusive transport transverse to the wires and does not localize.
One important difference between these constructions and the 2D hyperconductor is the diffusive, as opposed to ballistic, finite-temperature transport exhibited by the hyperconductor along the wires.

\acknowledgments

We thank S. Hartnoll, S.-S. Lee, S. Raghu, E. Shimshoni, and B. Ware for enjoyable and helpful discussions.
We also thank S. Hartnoll and S. Kivelson for comments on an early draft of the manuscript.
M.M. acknowledges the support of the John Templeton Foundation. C.N. has been partially supported by AFOSR under grant FA9550-10-1-0524.

\appendix

\section{Static Susceptibility Matrix}

\label{staticappendix}

The static susceptibility matrix $\hat{\chi}_{pq} = {1 \over L} G^R_{{\cal Q}_p {\cal Q}_q} (\omega = 0)$ where the conserved charges ${\cal Q}_{p}$ and ${\cal Q}_p$ of the action $S_{\rm b}$ involved in the retarded Green's function $G^R_{{\cal Q}_p {\cal Q}_q}$ are either one of the chiral electrical currents,
\begin{align} 
\label{currentcondensed}
J_I^e = & {e \over 2 \pi} {\rm sgn}(N - I) \int_x V_{IJ} \partial_x \phi_J \cr 
= &  {e \over 2 \pi} {\rm sgn}(N - I) \int_x V_{IJ} O_{J a} \partial_x X_a, 
\end{align}
or the Dirac momentum, 
\begin{align}
P_D = & -{1 \over 4 \pi} \int_x {\rm sgn}(N - I) \partial_x \phi_I \partial_x \phi_I \cr 
= & - {1 \over 4 \pi} {\rm sgn}(N - a) \int_x \partial_x X_a \partial_x X_a. 
\end{align}
In the above equations, $x \in (-L, L)$ with the length of the system $L \rightarrow \infty$, ${\rm sgn}(Z) = + 1$ for $Z \geq 0$ and ${\rm sgn}(Z) = - 1$ for $Z < 0$, and $J^e_I = J^e_{R,I}$ for $I = 1, \ldots, N$ and $J^e_I = J^e_{L, N - I}$ for $I = N + 1, \ldots, 2N$ with $N=23$.
Note that $I$ is not summed over on the right-hand side of Eq. (\ref{currentcondensed}).
We have introduced the fields $\phi_I = O_{I a} X_a$ with $O_{I a} \in SO(23,23)$ that diagonalize the action $S_{\rm b}$, tuned via the interaction matrix $V_{IJ}$ to the asymmetric Leech liquid point,
\begin{align}
S_{\rm b} = & {1 \over 4 \pi} \int_{t,x} \Big[{\rm sgn} (N - I) \partial_t \phi_I \partial_x \phi_I - V_{IJ} \partial_x \phi_I \partial_x \phi_J \Big] \cr
= & {1 \over 4 \pi} \int_{t,x} \Big[{\rm sgn} (N - a) \partial_t X_a \partial_x X_a - v \partial_x X_a \partial_x X_a \Big].
\end{align}
Henceforth, we set the velocity $v = 1$.
To isolate the leading temperature and frequency dependence of the conductivity, we need only compute the static susceptibility with respect to $S_{\rm b}$.

The bosonic action $S_{\rm b}$ enjoys the particle-hole symmetry $\phi_I \rightarrow - \phi_I$, $X_a \rightarrow - X_a$. 
Thus, the retarded Green's functions $G^R_{J^e_I P_D} = 0$ when computed with respect to $S_{\rm b}$ and so we focus upon the $J^e_I - J^e_J$ or $P_D - P_D$ static susceptibilities.
Scattering interactions at incommensurate fillings, interactions mediated by disorder, and higher-derivative band structure corrections to $S_{\rm b}$ generally break particle-hole symmetry and, thus, induce a non-zero overlap between the electrical currents and the momentum.
We ignore such overlaps as they contribute higher-order corrections to the conductivity than that to which we choose to work.
At commensurate fillings and in the absence of higher-derivative corrections, particle-hole symmetry is preserved.

To compute the retarded correlator, we exploit the relation $G^R_{{\cal Q}_p {\cal Q}_q} (\omega) = G^E_{{\cal Q}_p {\cal Q}_p} (i \omega_E \rightarrow \omega + i \delta)$ with the infinitesimal $\delta > 0$ between the retarded Green's function and the Euclidean Green's function at Euclidean  frequency $\omega_E$. 
The frequency $\omega$ of the retarded correlator has been analytically continued to the upper-half plane.
We shall often simply set $\delta = 0$ without mention.
Thus, the static susceptibility $\hat{\chi}_{p q} = {1 \over L} G^E_{{\cal Q}_p {\cal Q}_p} (\omega_E = 0)$.

We begin with the $J^e_I - J^e_J$ components of the static susceptibility,
\begin{align}
\label{staticcurrents}
\hat{\chi}_{J^e_I J^e_J} \equiv & {1 \over L} \lim_{\omega_E \rightarrow 0} \int_\tau e^{i \omega_E \tau} \Big\langle J_I^e(\tau) J_J^e(0) \Big\rangle \cr
= & {e^2 M^{ab}_{IJ} \over 4 \pi^2 L}  \lim_{\omega_E \rightarrow 0} \int_{\tau, x,y} e^{i \omega_E \tau} \Big\langle X'_a(\tau, x) X'_b(0, y) \Big\rangle, \cr
\end{align}
where $X'(\tau, x) \equiv \partial_x X(\tau, x)$,
\begin{align}
\label{currentcoefficient}
M_{IJ}^{a b} = & {\rm sgn}(N-I) {\rm sgn}(N-J) V_{IK} V_{J L} O_{Ka} O_{L b} \cr
= & {\rm sgn}(N-I) {\rm sgn}(N-J) (O^{-1})_{a I} (O^{-1})_{b J},
\end{align}
the Euclidean time $\tau \in [0, 1/T]$  and the brackets denote the thermal average at temperature $T$.
In simplifying Eq. (\ref{currentcoefficient}), we have made use of the identity $O_{I a} V_{IJ} O_{J b} = \delta_{ab}$.
Because $S_{\rm b}$ is diagonal when expressed in terms of the $X_a$ fields, the only non-zero correlators in Eq. (\ref{staticcurrents}) occur when $a=b$ and we obtain the well-known result,\cite{ginspargappliedcft1991}
\begin{align}
\label{currenttwopoint}
\langle X'_a(\tau, x) X'_b(0,0) \rangle = & - \delta_{ab} \Big({\pi T \over \sinh\Big(\pi T (x - {\rm sgn}_a i \tau) \Big)}\Big)^2,
\end{align}
where we have used the short-hand, ${\rm sgn}_a = {\rm sgn}(N - a)$.
It will be convenient to calculate a slightly more general Fourier transform than Eq. (\ref{staticcurrents}) by replacing the exponent in Eq. (\ref{currenttwopoint}), $2 \rightarrow 2 h$ with $h$ assumed to be half-integral.
Thus, we consider
\begin{widetext}
\begin{align}
\label{generalchiral}
{1 \over L} \int_{\tau, x, y} e^{i \omega_E \tau} \Big({\pi T \over \sinh\Big(\pi T (x-y - {\rm sgn}_a i \tau)\Big)}\Big)^{2 h} = & - {(\pi T)^{2 h} \over 2 L} \int_{x_+, x_-, \tau} e^{i \omega_E \tau} {1 \over \Big(\sinh\Big(\pi T (x_- - {\rm sgn}_a i \tau )\Big)^{2 h}} \cr
= & - \pi^{2 h} (2 T)^{2 h - 1} \int_{x_-} e^{{\rm sgn}_a 2 \pi T x_-}{1 \over 2 \pi i} \oint_{|\zeta| = 1} {\zeta^{{\omega_E \tau \over 2 \pi T} + h - 1} \over (\zeta - e^{{\rm sgn}_a 2 \pi T x_-})^{2 h}} \cr
= & - {T^{2 h - 1} \over 2 \omega_E} {(2 \pi)^{2 h} \over (2 h - 1)!} \prod_{i = 1}^{2 h - 1} \Big({\omega_E \over 2 \pi T} + h - i \Big).
\end{align}
\end{widetext}
In the first line, we made the change of variables, $x_{\pm} = x \pm y$ and then integrated over $x_+$; in the second line, we made the change of variable $\zeta = \exp(2 \pi T i \tau)$, performed the contour integration about the circle $|\zeta| = 1$, and then integrated over $x_-$.
Thus, we find for the current-current static susceptibility:
\begin{align}
\hat{\chi}_{J^e_I J^e_J} = & {e^2 \over 4 \pi} \sum_{a = 1}^{2N} M^{a a}_{IJ} \cr
= & {e^2 \over 4 \pi} {\rm sgn}(N-I) {\rm sgn}(N-J) V_{IJ},
\end{align}
where $I, J$ are not summed over and we used the relation $(O^{-1})^T.(O^{-1}) = V$.

Following an analogous procedure, we now calculate the $P_D-P_D$ static susceptibility,
\begin{align}
\hat{\chi}_{P_D P_D} \equiv & {1 \over L} \lim_{\omega_E \rightarrow 0} \int_\tau e^{i \omega_E \tau} \Big\langle P_D(\tau) P_D(0) \Big\rangle \cr
= & {2 \over 16 \pi^2 L} {\rm sgn}(N - a) {\rm sgn}(N - b) \cr
& \times  \int_{\tau, x, y} e^{i \omega_E \tau} \Big\langle X'_a(\tau, x) X'_b(0, y) \Big\rangle^2 \cr
= & {1 \over 8 \pi^2 L} \int_{\tau, x, y} e^{i \omega_E \tau} \Big\langle X'_a(\tau, x) X'_a(0, y) \Big\rangle^2, \cr
\end{align}
where we used Wick's theorem in going from the first to the second line and the fact that the only non-zero correlators occur when $a=b$ in going from the second to the third line.
We may now borrow the general result in Eq. (\ref{generalchiral}) by setting $h=2$ to conclude:
\begin{align}
\hat{\chi}_{P_D P_D} = {N \pi^2 T^2 \over 6}.
\end{align}

\section{Memory Matrix Elements}
\label{memorymatrixcomputations}

Recall the definition of the memory matrix reviewed Sec. \ref{memorymatrixsection} which we repeat here for convenience.
The memory matrix $\hat{M}(\omega)$ (the temperature dependence is left implicit) is defined as follows:
\begin{align}
\label{fullmemorymatrix}
\hat{M}(\omega) =& \sum_\alpha \Big(\lambda_\alpha^2 \hat{\cal M}^u_\alpha(\omega) + (\lambda^{\rm dis}_\alpha)^2 D \hat{\cal{M}}^{\rm dis}_\alpha(\omega)\Big), \\ 
\label{umklappelements}
(\hat{\cal M}^{\rm u})^{pq}_\alpha =& \frac{1}{L}\frac{\langle F^{\rm u}_{p, \alpha}; F^{\rm u}_{q, \alpha} \rangle_\omega - \langle F^{\rm u}_{p, \alpha}; F^{\rm u}_{q, \alpha} \rangle_{\omega=0} }{ i \omega}, \\
\label{disorderelements}
(\hat{\cal M}^{\rm dis})^{pq}_\alpha =& \frac{1}{L}\frac{\langle F^{\rm dis}_{p, \alpha}; F^{\rm dis}_{q, \alpha} \rangle_\omega - \langle F^{\rm dis}_{p, \alpha}; F^{\rm dis}_{q, \alpha} \rangle_{\omega=0} }{ i \omega}.
\end{align}
Here, $F^{\rm u}_{q, \alpha} = {i \over \lambda_\alpha} \left[H^{\rm u}_{\alpha}, {\cal Q}_q \right]$, $F^{\rm dis}_{q, \alpha} = {i \over \lambda^{\rm dis}_\alpha \sqrt{D}} \left[H^{\rm dis}_{\alpha}, {\cal Q}_q \right]$, and ${\cal Q}_q$ is a conserved charge (either $J^e_{R/L, I}$ or $P_D$). 
$\langle F^{\rm u}_{p, \alpha}; F^{\rm u}_{q, \alpha} \rangle_\omega$ and $\langle F^{\rm dis}_{p, \alpha}; F^{\rm dis}_{q, \alpha} \rangle_\omega$ are retarded finite-temperature Green's functions evaluated using $S_{\rm b}$.
$\lambda_\alpha$ and $\lambda^{\rm dis}_\alpha$ parameterize the umklapp scattering and coupling to disorder, respectively, and $D$ is the disorder variance of the Gaussian-correlated disorder, $\overline{\xi_\alpha(x)} = 0$, $\overline{\xi_\alpha(x) \xi_\beta^\ast(y)} = D \delta_{\alpha \beta}  \delta(x-y)$.
For simplicity, we take $\lambda_\alpha = \lambda$ and $\lambda_\alpha^{\rm dis} = \lambda^{\rm dis}$ for all $\alpha$.
$\hat{{\cal M}}^{\rm u}$ contains the contributions to the memory matrix from umklapp scattering, while $\hat{{\cal M}}^{\rm dis}$ contains the contributions from the disorder-mediated interaction. 
We stress that the form of the memory matrix given above is correct to leading order in the scattering interaction.
See Refs. [\onlinecite{Forster1975, Giamarchibook, Shimshoni2003, Hartnolllectures2013, LucasSachdev2015MM}] for further discussion.

\subsection{Evaluation of the $F^{\rm u, dis}_{p, \alpha}$}
\label{commutatorevaluations}

To compute the $F^{{\rm u}, {\rm dis}}_{p, \alpha}$ commutators, we make use of the equal-time commutators:
\begin{align}
\left[e^{i m_J^{(\alpha)} \phi_J(x)}, {\phi'_I(y) \over 2 \pi}\right] = m_I^{(\alpha)} {\rm sgn}_I \delta(x-y) e^{i m_J^{(\alpha)} \phi_J(x)}.
\end{align}
We find for the commutators $F^{\rm u}_{p, \alpha}$ of the ${\cal Q}_p$ with the umklapp scattering operators:
\begin{align}
\label{umklappcomms1}
F^{\rm u}_{J^e_I, \alpha} & = - 2 e\ {\rm sgn}(N-I) {\rm sgn}(N-J) V_{IJ} m_J^{(\alpha)} \cr
& \times \int_x {1 \over a^2} \sin\Big(\Delta k_{\alpha} x + m_K^{(\alpha)} \phi_K \Big), \\
\label{umklappcomms2}
F^{\rm u}_{P_D, \alpha} & = 2 \Delta k_\alpha \int_x {1 \over a^2} \sin\Big(\Delta k_{\alpha} x + m_K^{(\alpha)} \phi_K \Big), \cr
\end{align}
where the momentum mismatch $\Delta k_{\alpha} \equiv \sum_I m_I^{(\alpha)} k_F - p^{(\alpha)} G$, $G$ is a basis vector for the reciprocal lattice, and we have taken the Fermi momenta in all channels to be equal.
Recall that $a$ is a short-distance cutoff.
We see that the Dirac momentum $P_D$ commutes with the umklapp operators when $\Delta k_\alpha =0$, i.e., when the translation symmetry of the low-energy effective theory is preserved.
The result for $[H^{\rm u}_\alpha, P_D]$ is found using the integration by parts,
\begin{align}
\int_x e^{i \Delta k_\alpha x} {m_K^{(\alpha)} \over 2} \{\phi'_K, e^{i m_L^{(\alpha)} \phi_L} \}
& \equiv - i \int_x e^{i \Delta k_\alpha x} \partial_x e^{i m_L^{(\alpha)} \phi_L} \cr
& = - \Delta k_\alpha \int_x e^{i \Delta k_\alpha x + i m_L^{(\alpha)} \phi_L},
\end{align}
where we have dropped the boundary term and have defined the derivative operator on the right-hand side of the first line via a symmetric ordering prescription: $\partial_x \exp(i m^{(\alpha)}_I \phi_I) \equiv {i \over 2} m_J^{(\alpha)} \Big( \partial_x \phi_J \exp(i m_I^{(\alpha)} \phi_I) + \exp(i m_I^{(\alpha)} \phi_I) \partial_x \phi_J\Big)$.

The commutators $F^{{\rm dis}}_{p, \alpha}$ of the ${\cal Q}_p$ with the disorder-mediated interactions are computed in a similar fashion:
\begin{align}
\label{disordercomms1}
F^{\rm dis}_{J^e_I, \alpha} & = {i e \over \sqrt{D}} {\rm sgn}(N-I) {\rm sgn}(N-J) V_{IJ} m_J^{(\alpha)} \cr
& \times \int_x \Big[ \xi_\alpha(x) {1 \over a^2} e^{i m_K^{(\alpha)} \phi_K} - {\rm h.c.} \Big], \\ 
\label{disordercomms2}
F^{\rm dis}_{P_D, \alpha} & = - {1 \over \sqrt{D}} v^2 \int_x \Big[\Big(\partial_x \xi_\alpha(x)\Big) {1 \over a^2} e^{i m_K^{(\alpha)} \phi_K} + {\rm h.c.} \Big]. \cr
\end{align}
We see that the umklapp commutators in Eqs. (\ref{umklappcomms1}, \ref{umklappcomms2}) may be obtained from the disorder commutators in Eqs. (\ref{disordercomms1}, \ref{disordercomms2}) by substituting $\xi_\alpha(x) = \exp(i \Delta k_\alpha x)$.

\subsection{Evaluation of the $(\hat{\cal M}^{\rm u})^{pq}_\alpha$}
\label{umklappmemory}

We begin with the evaluation of the retarded two-point correlation functions $\langle F^{\rm u}_{p, \alpha}; F^{\rm u}_{q, \beta} \rangle_\omega$.
To leading order in the umklapp (and disorder) perturbations, these correlators are only non-zero when $\alpha = \beta$ because of the linear independence of the $m_I^{(\alpha)}$ so we set $\alpha = \beta$ in the remainder. 
Also, notice that $\langle F^{\rm u}_{p, \alpha}; F^{\rm dis}_{q, \beta} \rangle_\omega = 0$ because the disorder we study has zero mean, $\overline{\xi_\alpha(x)} = 0$. 
We simplify the following expressions by introducing the coefficients:
\begin{align}
\label{umklappcoefficients}
 U_{J^e_{I}, \alpha} = & - 2 e {\rm sgn}(N-I) {\rm sgn}(N-J) V_{IJ} m_J^{(\alpha)}, \cr
 U_{P_D, \alpha} = & 2 v^2 \Delta k_\alpha.
 \end{align}
 We see that $U_{P_D, \alpha} = 0$ for commensurate fillings when $\Delta k_\alpha = 0$ because translation invariance in the low-energy effective theory $S_{\rm lin}$ (interpreted as Dirac fermions created about zero-momentum) is preserved, resulting in divergent thermal conductivity.
 
Just as in Appendix \ref{staticappendix}, we compute the retarded correlators by Fourier transforming the Euclidean real-space correlation functions and then analytically continuing the Matsubara frequencies $\omega_E$ to real frequencies $\omega$ by way of the formula, $G^R_{F^{\rm u}_{p,\alpha} F^{\rm u}_{q, \alpha}}(\omega) = G^E_{F^{\rm u}_{p,\alpha} F^{\rm u}_{q, \alpha}}(i \omega_E \rightarrow \omega + i \delta) \equiv \langle F^{\rm u}_{p, \alpha}; F^{\rm u}_{q, \alpha} \rangle_{\omega_E \rightarrow - i \omega + \delta}$.

Thus, the Fourier transformed Euclidean correlation functions take the form:
\begin{widetext}
\begin{align}
\label{umklappcommcorrelators}
{1 \over L} \langle F^{\rm u}_{p, \alpha}; F^{\rm u}_{q, \alpha} \rangle_{\omega_E} = & {U_{p, \alpha} U_{q, \alpha} \over L} {1 \over a^4} \int_{x,y, \tau} e^{i \omega_E \tau} \Big\langle \sin\Big(\Delta k_\alpha x + m_K^{(\alpha)} \phi_K(\tau, x) \Big) \sin\Big(\Delta k_\alpha y + m_L^{(\alpha)} \phi_L(0, y) \Big) \Big\rangle \cr
= & {U_{p, \alpha} U_{q, \alpha} \over 4 L} \int_{x,y,\tau} e^{i \omega_E \tau} \Big[e^{i \Delta k_\alpha (x-y)} \Big\langle {e^{i m_K^{(\alpha)} \phi_K(\tau, x)} \over a^2} {e^{- i m_L^{(\alpha)} \phi_L(0, y)} \over a^2} \Big\rangle \cr
+& e^{- i \Delta k_\alpha (x-y)} \Big\langle {e^{- i m_K^{(\alpha)} \phi_K(\tau, x)} \over a^2} {e^{i m_L^{(\alpha)} \phi_L(0, y)} \over a^2} \Big\rangle \Big] \cr
= & {U_{p, \alpha} U_{q, \alpha} \over 2 L} \int_{x,y,\tau} e^{i \omega_E \tau} \cos\Big(\Delta k_\alpha (x-y) \Big) {(\pi T)^4 \over \sinh^3\Big(\pi T ((x-y) - i \tau) \Big) \sinh\Big(\pi T ((x-y) + i \tau) \Big)},\cr
\end{align}
\end{widetext}
where $x, y \in (- L, L)$ with $L \rightarrow \infty$ and $\tau \in [0, 1/T]$.
The first equality follows from direct substitution of Eqs. (\ref{umklappcomms1}, \ref{umklappcomms2}); for the second equality, we have only retained the non-zero terms in the product; for the third equality, we have used the standard thermal real-space Euclidean two-point function of a dimension $(\Delta_R, \Delta_L) = (3/2, 1/2)$ vertex operator ${1 \over \alpha^2} \exp(i m_J^{(\alpha)} \phi_J)$.\cite{ginspargappliedcft1991}
It is a great simplification of the calculation that all vertex operators considered have the same scaling dimension.
If only a fraction of the operators necessary to relax the currents had dimension $(3/2, 1/2)$ and the remaining operators were of higher dimension, it would be straightforward to calculate their effects by methods similar to those presented here. These operators would give subleading contributions to the memory matrix leading to slower relaxation of some conserved currents. As a result these operators would give the dominant contributions to the matrix of conductivities.

Similar to Appendix \ref{staticappendix}, we evaluate Eq. (\ref{umklappcommcorrelators}) by first making the change of variables $x_{\pm} = x \pm y$ and $\xi = e^{2 \pi i T \tau}$.
We assume a short-distance cutoff $0< a < |x - y|$.
The integral over $x_+$ factors out, canceling the $1/L$ prefactor, and we are left with the following integral to evaluate:
\begin{widetext}
\begin{align}
\label{umklappintegral}
{1 \over L} \langle F^{\rm u}_{p, \alpha}; F^{\rm u}_{q, \alpha} \rangle_{\omega_E} = & - 4 \pi^4 T^3 U_{p, \alpha} U_{q, \alpha} \int_{x_-} e^{- 2 \pi T x_-} \cos(\Delta k_\alpha x_-) {1 \over 2 \pi i} \oint_{|\zeta| = 1} {\zeta^{{\omega_E \over 2 \pi T} + 1} \over (\zeta - e^{- 2 \pi T x_-})^3 (\zeta - e^{2 \pi T x_-})} \cr
= & {\pi^2 T U_{p, \alpha} U_{q, \alpha} \over 4} \int_a^\infty dx_- {e^{- \omega_E x_-} \cos(\Delta k_\alpha x_- ) \over \sinh^3(2 \pi T x_- )} \Big[4 \pi^2 T^2 + \omega_E^2 \sinh^2(2 \pi T x_-) + \pi T \omega_E \sinh(4 \pi T x_- )\Big]. \cr
\end{align}
\end{widetext}
Next, we Wick rotate, $\omega_E \rightarrow - i \omega + \delta$, Eq. (\ref{umklappintegral}) to obtain the retarded Green's function,
\begin{widetext}
\begin{align}
\label{umklappintegral2}
{1 \over L} \langle F^{\rm u}_{p, \alpha}; F^{\rm u}_{q, \alpha} \rangle_{\omega} = & {\pi^2 T U_{p, \alpha} U_{q, \alpha} \over 4} \int_a^\infty dx_- {e^{- \delta x_- + i \omega x_-} \cos(\Delta k_\alpha x_- ) \over \sinh^3(2 \pi T x_- )} \Big[4 \pi^2 T^2 + (- i \omega + \delta)^2 \sinh^2(2 \pi T x_-) \cr
+ & \pi T (- i \omega + \delta) \sinh(4 \pi T x_- )\Big]. 
\end{align}
\end{widetext}

The remaining integral in Eq. (\ref{umklappintegral}) can be evaluated exactly to obtain the memory matrix elements $(\hat{\cal M}^{\rm u})^{pq}_\alpha$ defined in Eq. (\ref{umklappelements}).
The exact expression is rather complicated and so we shall examine it in various low-frequency and low-temperature limits for both commensurate and incommensurate fillings.
To study the low-frequency and low-temperature behavior of $(\hat{\cal M}^{\rm u})^{pq}_\alpha$, we first perform two expansions.
First, we expand the result as the short-distance cutoff $a \rightarrow 0$, keeping only the singular and finite non-zero terms.
Any $a \rightarrow 0$ singularities are a reflection of the short-distance divergences of the correlation function.
Second, we expand to linear order in $\delta$, however, we find it sufficient to study the resulting expression at $\delta = 0$ as the real part of the memory matrix is generally non-zero at finite $\omega$ and finite $T$.

\subsubsection{Commensurate Fillings}
\label{commensurateumklapp}

For commensurate fillings we set $\Delta k_\alpha = 0$.
For $\omega/T \ll 1$, the expression for the memory matrix element at commensurate fillings has the following behavior,
\begin{align}
\label{commensurateDC}
(\hat{\cal M}^{\rm u})^{pq}_\alpha\Big({\omega \over T} \ll 1\Big) = U_{p, \alpha} U_{q, \alpha} \Big[{\pi^4 \over 32} T + i {\pi \omega \over 16} \log(a_1 T)\Big],
\end{align}
where we have dropped all ${\cal O}(\delta)$ terms and absorbed all constants via a redefinition of the cutoff $a \rightarrow a_1$.
We shall make these multiplicative redefinitions of the short-distance cutoff $a \rightarrow a_i$ in each of the following expressions.
In the opposite regime when $T/\omega \ll 1$, we find the following expression for the memory matrix elements at commensurate filling,
\begin{align}
\label{commensurateAC}
(\hat{\cal M}^{\rm u})^{pq}_\alpha\Big({T \over \omega} \ll 1\Big) = U_{p, \alpha} U_{q, \alpha} \Big[ {\pi^2 \over 32} \omega + i {\pi \over 16} \omega \log(a_2 \omega) \Big],
\end{align}
where $a_1 \neq a_2$.

\subsubsection{Incommensurate Fillings}
\label{incommensurateumklapp}

When the filling is incommensurate, $\Delta k_\alpha \neq 0$. 
We shall study the memory matrix for frequencies and temperatures $\omega, T \ll \Delta k_\alpha$.

For $\omega/T \ll 1$, the expression for the memory matrix elements at incommensurate fillings have the following behavior,
\begin{align}
(\hat{\cal M}^{\rm u})^{pq}_\alpha\Big({\omega \over T} \ll 1\Big) = & U_{p, \alpha} U_{q, \alpha} \Big[{\pi^2 \over 32}\Big({(\Delta k_\alpha)^2 \over T} + 4 \pi^2 T \Big)e^{- {\Delta k_\alpha \over 2 T}} \cr 
+ &  i {\pi \omega \over 16} \log(a_3 \Delta k_\alpha) \Big],
\end{align}
where we have only retained the leading term present for $T \rightarrow 0$.
Precisely at $T=0$ (but first $\omega \rightarrow 0$), the real part of the $(\hat{\cal M}^{\rm u})^{pq}_\alpha\Big({\omega \over T} \ll 1\Big)$ vanishes when $\Delta k_\alpha \neq 0$ and we obtain a purely imaginary memory matrix which implies a finite Drude weight.
When $T/\omega \ll 1$, the incommensurate memory matrix takes the form,
\begin{align}
\label{incommensuratefrequencymemorymatrix}
(\hat{\cal M}^{\rm u})^{pq}_\alpha\Big({T \over \omega} \ll 1\Big) = & U_{p, \alpha} U_{q, \alpha} \Big[{\pi^2 \over 16} \Big({(\Delta k_\alpha)^2 \over \omega} + \omega \Big) e^{{\omega - \Delta k_\alpha \over 2 T}} \cr
+ & {i \pi \over 32}\Big(\omega \log\Big(a_4^2 ((\Delta k_\alpha)^2 - \omega^2)\Big) \cr 
+ & {(\Delta k_\alpha)^2 \over \omega} \log\Big(1 - {\omega^2 \over (\Delta k_\alpha)^2} \Big) \Big)\Big].
\end{align}

While we have studied the memory matrix for incommensurate fillings in the limit $\omega, T \ll \Delta k_\alpha$, we have checked that the initial expression obtained before taking the low-frequency or low-temperature limits reverts to the commensurate values by taking $\Delta k_\alpha = 0$.

\subsection{Evaluation of the $(\hat{\cal M}^{\rm dis})^{pq}_\alpha$}
\label{disordermemory}

Because the same vertex operators are used in both the umklapp and disorder-mediated interactions, the calculation of the disorder memory matrix elements $(\hat{\cal M}^{\rm dis})^{pq}_\alpha$ will be very similar to that of the previous section.
We begin with the evaluation of the retarded two-point correlation functions $\langle F^{\rm dis}_{p, \alpha}; F^{\rm dis}_{q, \alpha} \rangle_\omega$ which we determine by analytically continuing the Euclidean correlator $\langle F^{\rm dis}_{p, \alpha}; F^{\rm dis}_{q, \alpha} \rangle_{\omega_E}$.
We again simplify the ensuing expressions by introducing the coefficients,
\begin{align}
\label{disordercoefficients}
 \tilde{U}_{J^e_{I}, \alpha} = & i \, e \, {\rm sgn}(N-I) {\rm sgn}(N-J) V_{IJ} m_J^{(\alpha)}, \cr
 \tilde{U}_{P_D, \alpha} = & - v^2,
 \end{align}
 that occur in the disorder commutators in Eqs. (\ref{disordercomms1},\ref{disordercomms2}).

Unlike the correlators of the commutators involved in the umklapp calculation, we need to examine each set of correlators $\langle F^{\rm dis}_{J^e_I, \alpha}; F^{\rm dis}_{J^e_J, \alpha} \rangle_{\omega_E}$, $\langle F^{\rm dis}_{J^e_I, \alpha}; F^{\rm dis}_{P_D, \alpha} \rangle_{\omega_E}$, and $\langle F^{\rm dis}_{P_D, \alpha}; F^{\rm dis}_{P_D, \alpha} \rangle_{\omega_E}$ in turn.
First consider:
\begin{widetext}
\begin{align}
\label{disordercorrelator1}
{1 \over L} \langle F^{\rm dis}_{J^e_I, \alpha}; F^{\rm dis}_{J^e_J, \alpha} \rangle_{\omega_E = i \omega + \delta} = & - {(\pi T)^4 \tilde{U}_{J^e_I, \alpha} \tilde{U}_{J^e_J, \alpha} \over L D} \int_{x, y, \tau} e^{i \omega_E \tau} { \xi_\alpha(x) \xi^\ast_\alpha(y) + \xi^\ast_\alpha(x) \xi_\alpha(y) \over \sinh^3\Big(\pi T ((x-y) - i \tau) \Big) \sinh\Big(\pi T ((x-y) + i \tau) \Big)} \cr
= & {\pi^2 T \tilde{U}_{J^e_I, \alpha} \tilde{U}_{J^e_J, \alpha} \over 4 L D} \int_{x_+} \int_a^{\infty} d x_- {e^{(- \delta + i \omega) x_-} \over \sinh^3(2 \pi T x_- )} \Big[ \xi_\alpha(x) \xi^\ast_\alpha(y) + \xi^\ast_\alpha(x) \xi_\alpha(y) \Big] \cr
\times & \Big[4 \pi^2 T^2 + (- i \omega + \delta)^2 \sinh^2(2 \pi T x_-) + \pi T (- i \omega + \delta) \sinh(4 \pi T x_- )\Big],
\end{align}
\end{widetext}
where $x_{\pm} = x \pm y$ and we have performed identical manipulations to those explained in the previous section to evaluate Eqs. (\ref{umklappcommcorrelators}), (\ref{umklappintegral}), and (\ref{umklappintegral2}).

To explicitly evaluate the integrals over $x_+$ and $x_-$ in Eq. (\ref{disordercorrelator1}), we must choose a form for the functions $\xi_\alpha(x)$ parameterizing the disorder.
As we have discussed, we have chosen to consider zero-mean Gaussian-correlated disorder, $\overline{\xi(x)} = 0, \overline{\xi_{\alpha}(x) \xi^{\ast}_{\alpha}(y)} = D \delta(x-y)$.
To make contact with the pure system calculation of umklapp scattering at incommensurate fillings, we comment that this form of the disorder may be obtained by choosing a disorder potential, $\xi_{\alpha}(x) = \int_{\Delta p_\alpha} \tilde{\xi}(\Delta p_\alpha) e^{i \Delta p_{\alpha} x}$ with $\tilde{\xi}(\Delta p_\alpha) = 1$.
We see that incommensurate fillings can be understood as a particular disorder realization with $\tilde{\xi}(\Delta p_\alpha) = \delta(\Delta p_\alpha - \Delta k_\alpha)$.

Before integrating over $x_+$ and $x_-$ in Eq. (\ref{disordercorrelator1}), we first disorder average.
This allows us to again factor out the $x_+$ integral to cancel the $1/L$ prefactor and also to replace the product of disorder potentials $\xi_\alpha(x)$ inside the first brackets by $2 D\delta(x-y)$, where the $\delta(x-y)$ is understood to evaluate all terms containing $x_- = a$, the short-distance cutoff.
We find:
\begin{widetext}
\begin{align}
{1 \over L} \langle F^{\rm dis}_{J^e_I, \alpha}; F^{\rm dis}_{J^e_J, \alpha} \rangle_{\omega} = & {\pi^2 T \tilde{U}_{J^e_I, \alpha} \tilde{U}_{J^e_J, \alpha} \over 2} {e^{(- \delta + i \omega) a} \over \sinh^3(2 \pi T a)} \Big[4 \pi^2 T^2 + (- i \omega + \delta)^2 \sinh^2(2 \pi T a) + \pi T (- i \omega + \delta) \sinh(4 \pi T a) \Big].
\end{align}
\end{widetext}

Next, consider ${1 \over L} \langle F^{\rm dis}_{J^e_I, \alpha}; F^{\rm dis}_{P_D, \alpha} \rangle_{\omega}$.
The calculation of this correlator is identical to the previous one except that the overall coefficient now involves the $\tilde{U}_{J^e_I, \alpha} \tilde{U}_{P_D, \alpha}$ and the product of disorder potentials in the first line of Eq. (\ref{disordercorrelator1}) is replaced:
\begin{align}
\label{numeratorreplacement}
\xi_\alpha(x) \xi^\ast_\alpha(y) + \xi^\ast_\alpha(x) \xi_\alpha(y) \rightarrow & \xi_\alpha(x) \partial_y \xi^\ast_\alpha(y) - \xi^\ast_\alpha(x) \partial_y \xi_\alpha(y) \cr
= & \partial_y \Big(\xi_\alpha(x) \xi^\ast_\alpha(y) - \xi^\ast_\alpha(x) \xi_\alpha(y)\Big).
\end{align}
Upon disorder averaging, the term in the parentheses in Eq. (\ref{numeratorreplacement}) vanishes.
Thus, we find: 
\begin{align}
{1 \over L} \langle F^{\rm dis}_{J^e_I, \alpha}; F^{\rm dis}_{P_D, \alpha} \rangle_{\omega} = 0.
\end{align}
There is no overlap to leading order in the disorder-variance $D$ between the electrical and thermal currents.

Finally, we evaluate ${1 \over L} \langle F^{\rm dis}_{P_D, \alpha}; F^{\rm dis}_{P_D, \alpha} \rangle_{\omega}$ by replacing in Eq. (\ref{disordercorrelator1}):
\begin{align}
\tilde{U}_{J^e_I, \alpha} \tilde{U}_{J^e_J, \alpha} \rightarrow & \tilde{U}_{P_D, \alpha} \tilde{U}_{P_D, \alpha}, \cr
\xi_\alpha(x) \xi^\ast_\alpha(y) + \xi^\ast_\alpha(x) \xi_\alpha(y) \rightarrow & \partial_x \xi_\alpha(x) \partial_y \xi^\ast_\alpha(y) + {\rm h.c.} \cr
\end{align}
Disorder averaging, performing the integration by parts with respect to $\partial_{x/y} = \partial_{x_+} \pm \partial_{x_-}$, discarding all boundary terms, and evaluating $x_- = a$, we find:
\begin{widetext}
\begin{align}
{1 \over L} \langle F^{\rm dis}_{P_D, \alpha}; F^{\rm dis}_{P_D, \alpha} \rangle_{\omega} = & {\pi^2 T \tilde{U}_{P_D} \tilde{U}_{P_D} \over 2} \partial_{x_-} \partial_{x_-} \Big[{e^{(- \delta + i \omega) x_-} \over \sinh^3(2 \pi T x_-)} \Big[4 \pi^2 T^2 
+ (- i \omega + \delta)^2 \sinh^2(2 \pi T x_-) \cr
+ & \pi T (- i \omega + \delta) \sinh(4 \pi T x_-) \Big] \Big] \Big|_{x_- = a}
\end{align}
\end{widetext}

Equipped with the above correlation functions, we may now evaluate the memory matrix elements $(\hat{\cal M}^{\rm dis})^{J^e_{I} J^e_{I}}_\alpha$ and $(\hat{\cal M}^{\rm dis})^{P_D P_D}_\alpha$.
As before, we determine these memory matrix elements by expanding about the limit $a \rightarrow 0$ and subsequently expanding about $\delta = 0$.
It is sufficient to set $\delta = 0$.
In summary, we find:
\begin{widetext}
\begin{align}
\label{disorderelectricmemory}
(\hat{\cal M}^{\rm dis})^{J^e_{I} J^e_{I}}_\alpha = & \tilde{U}_{J^e_I, \alpha} \tilde{U}_{J^e_J, \alpha}\Big[ {2 \pi^3 \over 3} T^2 + {\pi \over 6} \omega^2 - i {3 \pi \over 24} {\omega \over a} \Big],\\
\label{disorderelectricmomentummemory}
(\hat{\cal M}^{\rm dis})^{J^e_{I} P_D}_\alpha = & 0, \\
\label{disordermomentummemory}
(\hat{\cal M}^{\rm dis})^{P_D P_D}_\alpha = & \tilde{U}_{P_D, \alpha} \tilde{U}_{P_D, \alpha}\Big[ {8 \pi^5 \over 5} T^4 + {2 \pi^3 \over 3} T^2 \omega^2 + {\pi \over 15} \omega^4 + i {\pi \over 15} {\omega \over a^3} \Big].
\end{align}
\end{widetext}

We notice that the logarithmic singularities that occurred in the umklapp memory matrix elements for $a=0$ are replaced by power-law singularities.
Such singularities reflect the short-distance divergences as correlation function insertion points become coincident.
They only occur in the imaginary part of the memory matrix elements at finite frequencies.
Our prescription is to remove such power-law divergences by hand and concentrate on the real parts of the memory matrix elements that determine the long-wavelength response of the system.
This prescription leads to agreement with related computations\cite{KaneFisher1992, FendleyLudwigSaleur1995} studying the tunneling conductance between quantum wires at a single point contact.

\section{$\hat{N}$ Matrix}
\label{Nmatrixappendix}

In this appendix, we show that $\hat{N} = 0$ to quadratic order in the umklapp $\lambda$ and disorder $\lambda^{\rm dis}$ couplings using rather general considerations.
Recall the definition:
\begin{align}
(\hat{N})_{pq} \equiv \hat{\chi}_{p \dot{q}} = \Big({\cal Q}_p, i [\sum_\alpha (H^{\rm u}_{\alpha} + H^{\rm dis}_\alpha), {\cal Q}_q])\Big).
\end{align}

\subsection{Umklapp Contributions}

First, consider the contribution to $\hat{N}$ from umklapp processes $H^{\rm u}_{\alpha}$.
Observe that $i [H^{\rm u}_\alpha, {\cal Q}_q] = \lambda F^{\rm u}_{q, \alpha}$ and $i [H^{\rm dis}_\alpha, {\cal Q}_q] = \sqrt{D} \lambda^{\rm dis}$, where ${\cal Q}_q \in \{J^e_I, P_D \}$, so that by using the definition of the static susceptibility and conventions in Appendix \ref{staticappendix}:
\begin{align}
(\hat{N})_{pq} = {\lambda \over L} \lim_{\omega_E \rightarrow 0} \int_\tau e^{i \omega_E \tau} \langle {\cal Q}_p(\tau) F^{\rm u}_{q, \alpha}(0) \rangle,
\end{align}
and likewise for the disorder contribution studied momentarily where the bracket denotes the Euclidean correlation function at temperature $T$.
At leading order in $\lambda$, the above two-point function $\langle {\cal Q}_p(\tau) F^{\rm u}_{q, \alpha}(0) \rangle$ vanishes when computed with respect to $S_{\rm b}$; more specifically, $\langle \partial_x \phi_I(\tau, x) e^{i m_J^{(\alpha)} \phi_J(0,y)} \rangle = 0$ and $\langle \partial_x \phi_I(\tau, x)  \partial_x \phi_I(\tau, x) e^{i m_J^{(\alpha)} \phi_J(0,y)} \rangle = 0$ when computed with respect to $S_{\rm b}$.
At quadratic order, $\lambda^2$, there is the correction,
\begin{align}
\label{quadraticumklapp}
\delta (\hat{N})_{pq} = {\lambda^2 \over L} \lim_{\omega_E \rightarrow 0} \int_{\tau, \tau', z} e^{i \omega_E \tau} \langle {\cal Q}_p(\tau) F^{\rm u}_{q, \alpha}(0) H^{\rm u}_{\alpha}(\tau', z) \rangle.
\end{align}
The above correlation function, computed with respect to $S_{\rm b}$ factorizes, into the sum of two three-point functions:
\begin{widetext}
\begin{align}
\label{3pointfunction}
{\lambda^2 \over L} \int_{\tau, \tau', z} e^{i \omega_E \tau} \langle {\cal Q}_p(\tau) F^{\rm u}_{q, \alpha}(0) H^{\rm u}_{\alpha}(\tau', z) \rangle \propto & {i \lambda^2 (\pi T)^5 \over L} \int_{\tau, \tau', x, y, z} e^{i \omega_E \tau} \Big[{C_1 e^{- i \Delta k_\alpha X_{zy}} - C_2 e^{i \Delta k_\alpha X_{zy}} \over \sinh\Big(\pi T(X_{zy} + i \tau') \Big)} \Big] \cr
\times & {1 \over \sinh^h\Big(\pi T(X_{xy} - i \tau ) \Big) \sinh^h\Big(\pi T(X_{xz} - i \tau +i \tau' ) \Big) \sinh^{3 - h}\Big(\pi T(X_{zy} - i \tau' ) \Big)}, \cr
\end{align}
\end{widetext}
for constants $C_1 = (-1)^h C_2$ (whose precise magnitude will not be required) equal to the operator product coefficients for the fusion, ${\cal Q}_p \exp(i m_I^{(\alpha)} \phi_I) \sim \exp(i m_I^{(\alpha)} \phi_I)$, and $h=1$ when ${\cal Q}_p = J^e_I$ and $h=2$ when ${\cal Q}_p = P_D$.
Above, we have introduced the ``difference coordinates" $X_{xy} = x-y, X_{xz} = x-z, X_{zy} = z-y$.
At $\omega_E = 0$, we notice that the integrand is odd under the reflection of all spatial and temporal coordinates followed by the shifts, $\tau, \tau' \rightarrow \tau - 1/T, \tau' - 1/T$.
Therefore, the integral is zero at $\omega_E=0$ and the quadratic contribution to $\hat{N}$ from umklapp processes vanishes.

\subsection{Disorder Contributions}

Next, consider the contributions to $\hat{N}$ from disorder-mediated processes $H^{\rm dis}_\alpha$.
The term linear in $\lambda^{\rm dis}$ again vanishes for the same reason as before.
At quadratic order, we consider Eq. (\ref{quadraticumklapp}) with the superscript ${\rm u}$ replaced by ${\rm dis}$.
The form of the resulting three-point function is very similar to that which appears in Eq. (\ref{3pointfunction}).
The difference is due to the disorder $\xi_\alpha$ appearing in the disorder commutators Eqs. (\ref{disordercomms1}, \ref{disordercomms2}) and $H^{\rm dis}_\alpha$.
For $F^{\rm dis}_{q, \alpha} = F^{\rm dis}_{J^e_I, \alpha}$, we disorder average and insert $\delta(y-z)$ into integrand in Eq. (\ref{3pointfunction}) at $\omega_E = 0$: when ${\cal Q}_p = J^e_I$, the three-point function vanishes using the above reflection and translation argument; when ${\cal Q}_p = P_D$, the three-point function vanishes identically after setting $y=z$ and using $C_1 = C_2$ for $h=2$.
For $F^{\rm dis}_{q,\alpha} = F^{\rm dis}_{P_D, I}$, we disorder average, replace the relative minus sign between $C_1$ and $C_2$ by $(+1)$, and insert $\partial_y \delta(y-z)$ into the integrand in Eq. (\ref{3pointfunction}): when ${\cal Q}_p = J^e_I$, the integrand vanishes identically similar to $P_D$ before; when ${\cal Q}_p = P_D$, we may again apply the reflection and translation argument to conclude that the integral vanishes at $\omega_E = 0$.
Thus, we may safely ignore the $\hat{N}$ matrix in our transport calculations.

\section{Exact Marginality Along the `Decoupled Surface'}
\label{marginalityappendix}

In this Appendix, we argue perturbatively for the exact marginality, along the decoupled surface, of the dimension $(\Delta_R, \Delta_L) = (3/2, 1/2)$ operators used to relax the electrical and thermal currents.
Our argument strictly applies in the scaling limit in which only classically marginal and relevant interactions are retained in the low-energy effective theory with irrelevant interactions being set to zero.

Recall from Sec. \ref{PMreview} that the decoupled surface is a subspace within the hyperconductor phase in which the interaction matrix $\tilde{V}_{IJ}$ is block diagonal.
The scaling dimensions of operators are independent of $\tilde{V}_{IJ}$ when the theory lies on the decoupled surface; however, scaling dimensions vary continuously with $\tilde{V}_{IJ}$ upon departing from the decoupled surface.

We consider the collection of operators ${\cal O}_\alpha = \cos\Big(m_I^{(\alpha)} \phi_I\Big)$ with scaling dimension equal to $(3/2,1/2)$ along the decoupled surface whose coupling constants we denote by $g_\alpha$ . 
These operators are exactly marginal if their beta function $\beta_{g_\alpha}$ vanishes to all orders in the couplings of the theory,
\begin{align}
\label{RGequation}
\dot{g}_\alpha = \beta_{g_\alpha},
\end{align} 
where the dot denotes a variation of the coupling with respect to the renormalization group scale.
We will understand the contributions to $\beta_{g_\alpha}$ as arising from corrections to scaling (i.e., conformal perturbation theory) of the zero-temperature two-point function,
\begin{align}
\langle {\cal O}_\alpha(z, \bar{z}) {\cal O}_\alpha(0) \rangle \sim z^{-1} \bar{z}^{-3},
\end{align}
for $z = x + i \tau, \bar{z} = x - i \tau$ computed with respect to the fixed point action $S_{\rm b}$ in Eq. (\ref{eqn:LL-action}).\cite{ginspargappliedcft1991}

First, observe that ${\cal O}_{\alpha}$ has unit spin, $\Delta_R - \Delta_L$, under the $SO(2) = U(1)$ rotation symmetry of the Euclidean theory.
When the action is perturbed, $S_{\rm b} \rightarrow S_{\rm b} + g_\alpha \int {\cal O}_{\alpha}$, the $SO(2)$ symmetry is broken.  
We may view $g_\alpha$ as a spurion -- a ``field" that transforms oppositely to the operator it multiplies so that the product is an $SO(2)$ singlet -- of this broken rotational symmetry.
This means that $g_{\alpha}$ may be understood to have spin-(-1).
With this understanding, we may constrain the form of $\beta_{g_\alpha}$.

The left-hand side of Eq. (\ref{RGequation}) is linear in $g_\alpha$ and so the equality implies that $\beta_{g_\alpha}$ also carries spin-(-1).
Thus, we must determine what spin-1 combination of operators could possibly contribute to $\beta_{g_\alpha}$.\cite{Cardypotts1993}
Working in the scaling limit where all irrelevant operators are ignored allows us to disregard any contribution from high-dimension operators with negative spin.
There are no marginal spin-(-1) operators because the lowest scaling dimension of a right-moving vertex operator is equal to $3/2$.
There do exist spin-(-1) relevant and spin-(-2) marginal operators which are quadratic and quartic in the  fermions of the left-moving sector along with marginal spin-0, i.e., dimension $(1,1)$ operators, and spin-2 operators in addition to the marginal ${\cal O}_\alpha$ operators.
Perturbations by spin-(-1) operators can be absorbed by a field redefinition of the left-moving fermion sector and so we ignore such deformations.

A general contribution to the ${\cal O}_\alpha$ two-point function contains $N_{-2}$ spin-(-2) insertions, $N_0$ spin-0 insertions, $N_2$ spin-2 insertions, and $N_{{\cal O}_\beta}$ ${\cal O}_{\beta}$ insertions.
Note that we are collectively referring to all additional insertions of the ${\cal O}_\beta$ operators as $N_{{\cal O}_\beta}$. 
In order for $\beta_{{\cal O}_\alpha}$ to carry spin equal to -1, we require the number of insertions of various operators to satisfy:
\begin{align}
\label{spinconstraint}
2 N_{-2} - N_{{\cal O}_\beta} - 2 N_2 = -1.
\end{align}
Thus, $N_{{\cal O}_\beta}$ should be odd.

All operators in the left-moving sector can be built from products of the fermion operators and their spatial derivatives.
Since the left-moving sector is describable in terms of interacting chiral fermions, fermion parity constrains any non-zero contribution to the ${\cal O}_\alpha$ two-point function to contain an even number of left-moving fermion operators:
\begin{align}
\label{Wickconstraint}
4 N_{-2} + N_{{\cal O}_\beta} + 2 N_2 + 2 N_0 \in 2 \mathbb{Z}.
\end{align}
The first contribution to the left-hand side of Eq. (\ref{Wickconstraint}) assumes an operator quartic in the fermion operators. 
An operator that is only quadratic with a single spatial derivative acting on one of the fermions might also contribute. 
However, this has no effect on the conclusion that the parity of the left-hand side must be even.

Eqs. (\ref{spinconstraint}) and (\ref{Wickconstraint}) are not consistent with one another: the former requires $N_{{\cal O}_\beta}$ to be odd, while the latter requires $N_{{\cal O}_\beta}$ to be even.
The only resolution is that the ${\cal O}_\alpha$ operators are exactly marginal in the scaling limit and so $\beta_{g_\alpha} = 0$.
There is likewise no renormalization of the Luttinger liquid parameters of $S_{\rm b}$ due to the spin-1 ${\cal O}_\alpha$ operators.

Exact marginality of the dimension $(3/2, 1/2)$ operators and the Luttinger parameters along the decoupled surface is a consequence of the chirality or spin-1 nature of the ${\cal O}_{\alpha}$ operators which is ultimately due to the asymmetric nature of the left-moving and right-moving excitations in the asymmetric shorter Leech liquid underlying the hyperconductor studied in this paper.
The de-coupled renormalization group equations described above should be contrasted with those of the Kosterlitz-Thouless transition that involve a dimension $(1,1)$ vertex operator and the Luttinger parameter.\cite{Giamarchibook}
It is this difference that results in the logarithmic corrections to scaling in the expressions for the conductivities in the work of Giamarchi on transport in a 1D Luttinger liquid.\cite{Giamarchi1991}

\settocdepth{section}
\bibliography{PMTransport}

\end{document}